\documentclass[
 unsortedaddress,
 amsmath,amssymb,
 aps,
floatfix,
pra,
reprint,
prb,
]{revtex4-2}
\usepackage{caption}
\pdfoutput=1
\usepackage{subcaption}
\usepackage{graphicx}
\usepackage{dcolumn}
\usepackage{bm}
\usepackage{float}
\usepackage{amsmath,amssymb}
\usepackage{romannum}
\usepackage{multirow} 

\usepackage{cleveref}
\usepackage{booktabs} 
\usepackage{tablefootnote}
\usepackage{xcolor}
\usepackage{tikz}
\usetikzlibrary{arrows.meta, positioning, shapes.geometric}

\begin{document}

\preprint{APS/123-QED}

\title{Data-Efficient Training of Linear ACE Potentials through Leverage-Guided Subset Selection of ASSYST Structure Pools}

\author{Aynour Khosravi}
\email{Corresponding Author: aynour.khosravi@ubc.ca} 
\affiliation{Department of Materials Engineering, The University of British Columbia, Vancouver, Canada}
\author{Marvin Poul}
\email{poul@mpie.de}
\affiliation{Department of Computational Materials Design, Max-Planck-Institut für Eisenforschung GmbH, Düsseldorf, Germany}
\author{Jörg Neugebauer}
\email{neugebauer@mpi-susmat.de}
\affiliation{Department of Computational Materials Design, Max-Planck-Institut für Eisenforschung GmbH, Düsseldorf, Germany}
\author{Chad W. Sinclair}
\email{chad.sinclair@ubc.ca}
\affiliation{Department of Materials Engineering, The University of British Columbia, Vancouver, Canada}

\begin{abstract}
The construction of machine-learned interatomic potentials (MLIPs) is often limited by the cost of generating large density-functional-theory (DFT) training datasets. For systematically generated structure pools such as ASSYST, a central practical question is how many configurations must be labeled to achieve reliable accuracy. Here we assess geometry-based, label-free subset selection for training linear Atomic Cluster Expansion (ACE) potentials. Using statistical leverage scores and CUR-type sampling, we compare leverage-guided selection against random, energy-based, and force-based baselines under controlled iterative protocols. Elemental Al provides the primary benchmark, with Cu and Al--Cu alloys used for transfer validation. Leverage-guided subsets recover plateau-level energy and force accuracy using substantially smaller labeled fractions ($\sim$30--40\%) than random sampling, corresponding to an effective 2–3× reduction in DFT labeling for the systems studied. In alloy tests, defect energetics remain comparable across strategies once sufficient chemical diversity is included, while leverage selection maintains competitive accuracy at reduced training size.

These results demonstrate that descriptor-space–guided, label-free subsampling can significantly reduce DFT workload for linear ACE models trained on ASSYST structure pools without degrading defect-level fidelity.
\end{abstract}

\maketitle
\pagenumbering{arabic}

\section{Introduction}

Machine-learned interatomic potentials (MLIPs) provide parametric representations of the potential energy surface (PES), trained to reproduce quantum-mechanical energies and forces at a computational cost comparable to classical force fields~\cite{Mueller2020-tp,Behler2017,Deringer2019,No2020,Behler2016,Unke2021,Friederich2021}. Over the past decade, MLIPs have enabled large-scale atomistic simulations at system sizes and time scales inaccessible to direct first-principles calculations. Early developments include the Behler–Parrinello neural-network potential~\cite{Behler2007} and the Gaussian Approximation Potential~\cite{Bartk2010}. Subsequent frameworks encompass Moment-Tensor Potentials~\cite{Shapeev2016,Novikov2021}, the Atomic Cluster Expansion (ACE)~\cite{Drautz2019,Dusson2022}, Spectral Neighbour Analysis Potentials~\cite{Thompson2015,Wood2018}, message-passing neural networks~\cite{NEURIPS2022_4a36c3c5,Unke2019,Zubatyuk2019,Batzner2022,Schtt2021EquivariantMP}, and other deep neural architectures~\cite{Zhang2018,Wang2018}.

A central challenge in atomistic modelling is achieving electronic-structure accuracy while accessing the length and time scales relevant to materials behaviour. Density-functional theory (DFT) provides reliable access to the potential energy surface but becomes computationally prohibitive for large systems or long dynamical trajectories~\cite{Mohr2015}. Classical empirical force fields, such as embedded-atom method (EAM) potentials~\cite{Daw1984} offer excellent computational efficiency but often lack transferability in environments involving defects, significant lattice distortions, anharmonicity, or chemical disorder~\cite{OED_MLIP_2025}. MLIPs aim to bridge this gap by retaining near-classical efficiency while systematically approximating quantum-mechanical energetics and forces.

From a modeling perspective, constructing an MLIP corresponds to a high-dimensional regression problem over atomic configurations. Rather than operating directly in the full coordinate space $\mathbb{R}^{3N}$, local atomic environments are mapped onto symmetry-invariant descriptors, and regression models are trained to predict energies and forces. Descriptor complexity must be carefully balanced: representations that are too simple fail to distinguish physically distinct environments, whereas overly rich representations demand large training datasets and become more difficult to fit robustly~\cite{Blank1995,Bartk2010,OED_MLIP_2025}.

In this work, we focus on \emph{linear} ACE models~\cite{Drautz2019,Dusson2022}. ACE provides a symmetry-adapted and systematically improvable representation of local atomic environments while remaining linear in the model coefficients. This linear structure makes the regression problem explicit and analytically tractable, offering a transparent setting in which to examine the relationship between basis complexity and training-set design. In particular, it enables subset-selection strategies that depend only on geometric properties of the ACE design matrix, without requiring access to quantum-mechanical labels.

Regardless of model architecture, MLIPs are trained on quantum-mechanical reference data, most commonly DFT energies and forces~\cite{OptData_MLIP_2022}. For new material systems, the dominant computational cost in expanding a dataset typically arises from evaluating these quantum-mechanical labels rather than generating candidate atomic geometries~\cite{BayesSelection_2025,OptData_MLIP_2022}. In practice, training sets are assembled from molecular-dynamics snapshots, distorted structures, and selected defect or surface configurations. Such procedures can introduce substantial redundancy in descriptor space, while rare but physically significant environments remain sparsely sampled. This leads to a fundamental question: \emph{how much, and what kind, of quantum-mechanical data is necessary to achieve accurate and transferable MLIPs?}

Efficient dataset construction has consequently received significant attention in MLIP development~\cite{BayesSelection_2025}. Starting from a pool of candidate configurations, the task is to identify subsets whose quantum-mechanical labeling yields accurate potentials within constrained computational budgets. Such candidate pools may be generated through molecular dynamics with existing potentials, random atomic or cell perturbations, chemical substitutions, or curated defect and surface structures~\cite{OptData_MLIP_2022}. Recent work has also introduced automated and systematic frameworks for exploring potential-energy surfaces and constructing large, diverse training datasets, aiming to reduce the manual effort in dataset generation while improving transferability~\cite{Liu2025}\cite{pa2025information}\cite{fletcher2025autonomous}. Recent studies have also examined the joint optimization of DFT convergence settings, training-set size, and model complexity to characterize accuracy–cost trade-offs in linear MLIPs~\cite{BAGHISHOV2025332}.

Online active-learning strategies iteratively select configurations based on model uncertainty or committee disagreement and retrain the potential as new labels become available~\cite{AL_Leverage_2018}. While often effective, such workflows require repeated retraining and close integration with high-performance computing resources, and early iterations may be influenced by sampling bias when the model remains weakly constrained. An alternative viewpoint treats dataset construction as an \emph{offline} subset-selection problem, in which configurations are chosen to span descriptor space prior to expensive quantum-mechanical labelling~\cite{PCovR_CUR_2021,BayesSelection_2025}. In linear models, statistical leverage scores provide a natural measure of the influence of candidate configurations on the regression fit and offer principled criteria for prioritizing configurations for labeling~\cite{AL_Leverage_2018,OED_MLIP_2025}.

For multicomponent alloys, training-set construction is particularly challenging because both structural and compositional degrees of freedom must be sampled across a broad region of phase space. 
The Automated Small SYmmetric Structure Training (ASSYST) framework provides a systematic and material-agnostic approach for generating diverse candidate structures by sampling random crystal prototypes across space groups~\cite{Poul2023,Poul2025}. 
By focusing on small periodic cells (typically on the order of $\sim$10 atoms), ASSYST facilitates high-throughput DFT labeling while exposing a wide variety of local atomic environments through random generation, relaxation, and controlled perturbations. 
Potentials trained on sufficiently large ASSYST datasets have been shown to reproduce bulk phases, solid solutions, and defect energetics beyond the specific configurations included in the training set, and to yield phase diagrams in reasonable agreement with experiment~\cite{Poul2025}.

While ASSYST provides a systematic and transferable framework for generating large structure pools, it does not prescribe how these pools should be utilized for efficient model training. In particular, the number of configurations that must be labeled with high-accuracy DFT to achieve a target accuracy for a given model class remains unclear. Since ASSYST datasets may contain tens of thousands of structures, exhaustive labeling can be computationally demanding. This raises the central question of whether substantially smaller, carefully selected subsets can retain the predictive performance and transferability of the full dataset. In this work, we address this question using descriptor-based subset-selection strategies for linear ACE models trained on ASSYST-generated configurations.

Previous work on data efficient selection schemes for the construction of MLIP has not investigated in detail their impact on the transferability properties of the resulting potentials.
In~\cite{BayesSelection_2025} the selection starts from a manually created dataset and evaluates the performance of the schemes on a specific application target.
For elastic and vibrational properties, but only those, Ref.~\cite{OptData_MLIP_2022} demonstrates large efficiency gains, though by means of constructing the dataset rather than subsetting a candidate set. In
\cite{BAGHISHOV2025332} the authors used very diverse datasets, but tested the impact of data selection only on general fitting metrics, leaving the question open whether their schemes impacted the transferability of the final potentials.
While these results make it clear that very efficient databases may be constructed and ASSYST clearly shows that sufficiently diverse databases yield transferable potentials, it is not yet clear whether the combination is achievable in practice.
We focus here explicitly on this question---whether these schemes are applicable to general potentials or whether they achieve sparsity in training data by reducing transferability, i.e. collapsing the region of
descriptor space that must be sampled to achieve accuracy.

In this work, we examine this question for linear ACE models by formulating training-set design as a subset-selection problem on the ACE design matrix constructed from general ASSYST-generated configurations. Leveraging the explicit linear structure of ACE, we compute regularized leverage scores and apply CUR and block-CUR subsampling to identify configurations that are most informative for the regression, without using energy or force labels. Using elemental Al as the primary benchmark, with Cu and Al--Cu alloys as transfer tests, we compare leverage-based selection with random sampling and label-driven heuristics across labeled fractions and ACE basis settings.
We find that leverage-guided subsets containing approximately $30$--$40\%$ of the labeled ASSYST configurations achieve, within statistical uncertainty, the same predictive accuracy as models trained on the full datasets, while exhibiting smoother and more reproducible learning behaviour across bulk, defect, and alloy test cases. These results demonstrate that, for linear ACE models trained on ASSYST pools, substantial reductions in DFT labeling can be achieved without sacrificing predictive accuracy across physically distinct regimes, rather than merely maintaining accuracy on a fixed test set.

The paper is organized as follows. Section~\ref{sec:methodology} describes the computational setup: the generation of ASSYST pools (\ref{subsec:assyst_workflow}), the construction of the linear ACE basis and design matrix (\ref{subsec:ace_basis_design_matrix}), and the leverage-based CUR and block-CUR selection strategies (\ref{subsec:cur_blockcur}). Section~\ref{sec:results} presents the results: learning curves and data efficiency on Al (\ref{subsec:res_al_efficiency}), sensitivity to ACE hyperparameters (\ref{subsec:res_hyperparam_sensitivity}), transfer tests on Cu and Al--Cu (\ref{subsec:res_transfer_cu_alcu}), validation against independent DFT data (\ref{subsec:res_dft_validation}), and a summary of practical DFT savings (\ref{subsec:data_efficiency}). Section~\ref{section:Conclusion} concludes with implications for scalable training-set design and future directions.

\section{Methodology}
\label{sec:methodology}

\subsection{ASSYST Data Generation and Workflow Overview}
\label{subsec:assyst_workflow}

In this work, we use the ASSYST configuration pools for Al, Cu, and Al--Cu reported in Ref.~\cite{Poul2025,Poul2023}. These pre-generated pools serve as the starting point for our subset-selection and ACE fitting pipeline. ASSYST generates small periodic structures ($\sim$8--10 atoms) by systematically sampling random crystal prototypes across space groups, relaxing them with low-cost DFT, and applying controlled random perturbations to the atomic positions and lattice parameters. This procedure efficiently explores a broad range of local environments while requiring minimal user input and modest computational cost.

In our workflow, ASSYST provides a rich and unbiased sampling of configurational space, from which we subsequently select a compact and informative subset using leverage scores (see Sec.~\ref{subsec:leverage_scores}).  Table~\ref{tab:assystparameter} summarizes the parameters for ASSYST used in this study.

\begin{table}
    \centering
    \caption{\
        ASSYST input parameters for the three generation stages as described in detail in~\cite{Poul2025}.
        In total 9236 structures were generated for Al, 10631 for Cu, and 21330 for Al-Cu.
    }
    \label{tab:assystparameter} 
    \begin{tabular}{ccl|c}
        \toprule
        \multicolumn{2}{c}{Step}   & parameter & value      \\
        \midrule
        Random     & unary  & \#Atoms & 1--10 per element         \\
        Spacegroup & binary & \#Atoms & 1,2,3,4,6,8 per element \\
        Sampling   &        &         & total per cell $\leq 10$             \\
        \midrule
        \multicolumn{2}{c}{Vibrational} & $\sigma_\mathrm{rattle}$  & 0.25\,\AA \\
                    &                   & $n_\mathrm{rattle}$       & 4        \\
        \midrule
        \multicolumn{2}{c}{Elastic} & $\epsilon_\mathrm{hydro}$ & 0.6 \\
                        &           & $\epsilon_\mathrm{shear}$ & 0.15 \\
                        &           & $n_\mathrm{stretch}$      & 4   \\
        \bottomrule
    \end{tabular}
\end{table}

\subsection{ACE basis construction and design matrix}
\label{subsec:ace_basis_design_matrix}

In this work we employ the linear Atomic Cluster Expansion framework to parameterize the potential energy surface of Al, Cu and Al--Cu~\cite{Drautz2019, Witt2023, bochkarev2023acepotentials}. For a configuration with atomic positions $\{\mathbf r_i\}$ and chemical species $\{Z_i\}$, the total energy is decomposed into site energies,
\begin{equation}
  E = \sum_i \varepsilon_i,
  \label{eq:ace_total_energy}
\end{equation}
where each site contribution $\varepsilon_i$ depends only on the local neighbourhood of atom $i$ within a cutoff radius $r_{\mathrm{cut}}$ and respects translational, rotational, reflection, and permutation symmetries~\cite{Drautz2019}.

It provides a systematic, body-ordered expansion of $\varepsilon_i$ in terms of symmetric polynomials built from one-particle basis functions. Introducing $x_i := (\mathbf r_i, Z_i)$ for the state of atom $i$ and $x_{ij} := (\mathbf r_j - \mathbf r_i, Z_i, Z_j)$ for a bond, the canonical cluster expansion reads
\begin{align}
  \varepsilon_i
    &= V^{(0)}(Z_i)
     + \sum_{j_1} V^{(1)}(x_{ij_1})
     + \sum_{j_1<j_2} V^{(2)}(x_{ij_1},x_{ij_2})
     \notag \\
    &\quad
     + \cdots
     + \sum_{j_1<\cdots<j_{\bar\nu}}
       V^{(\bar\nu)}(x_{ij_1},\dots,x_{ij_{\bar\nu}}),
  \label{eq:canonical_expansion}
\end{align}
which is truncated at a maximum correlation order $\bar\nu$. In practice we use the equivalent ``self-interacting’’ formulation introduced in Ref.~\cite{Drautz2019,Witt2023}, which permits a tensor-product structure and efficient evaluation but can be transformed back to the canonical form if needed.

The angular dependence of the atomic neighbourhood is represented through spherical harmonics $Y_l^m$, while the radial dependence is expanded in a set of smooth radial basis functions. A general one-particle basis function centered on atom $i$ is written as
\begin{equation}
  \phi_{z n l m}(r_{ij}, Z_i, Z_j)
    = R_{n}(r_{ij}, Z_i, Z_j)\,
      Y_l^m(\widehat{\mathbf r}_{ij})\,
      \delta_{z Z_j},
  \label{eq:one_particle_basis}
\end{equation}
where $r_{ij} = |\mathbf r_j - \mathbf r_i|$, $\widehat{\mathbf r}_{ij} = (\mathbf r_j - \mathbf r_i)/r_{ij}$, and the Kronecker symbol $\delta_{z Z_j}$ selects the chemical species of the neighbour atom $j$.

Following the implementation in \textsc{ACEpotentials.jl}, the radial basis functions are indexed by $n$ only, i.e.\ $R_{nl} \equiv R_n$ for all $l$. Each radial function is constructed as~\cite{Witt2023}
\begin{equation}
  R_{n}(r_{ij}, Z_i, Z_j)
    = f_{\mathrm{env}}(r_{ij}, Z_i, Z_j)\,
      P_n\!\bigl(y(r_{ij}, Z_i, Z_j)\bigr),
  \label{eq:radial_basis}
\end{equation}
where $y(r)$ is a smooth, element-dependent distance transform designed to enhance resolution near typical bond lengths, $P_n$ is an orthogonal polynomial basis in the transformed coordinate (Legendre polynomials in this work), and $f_{\mathrm{env}}$ is a smooth envelope enforcing the cutoff radius $r_{\mathrm{cut}}$ and ensuring regular behaviour at both small and large interatomic separations~\cite{bochkarev2023acepotentials}.

Many-body contributions are obtained by forming tensor products of the one-particle basis. For a fixed central species $Z_i$, a $\nu$-body potential is expressed as
\begin{equation}
  V^{(\nu)}(x_{ij_1},\dots,x_{ij_{\nu}})
   = \sum_{k_1,\dots,k_{\nu}}
     c^{(Z_i)}_{k_1\cdots k_{\nu}}\,
     \phi_{k_1}(x_{ij_1}) \cdots \phi_{k_\nu}(x_{ij_\nu}),
  \label{eq:many_body_expansion}
\end{equation}
where $k_t = (z_t,n_t,l_t,m_t)$ indexes the underlying one-particle basis. After symmetry reduction, the expansion yields a linear representation of the site energy,
\begin{equation}
  \varepsilon_i = \mathbf c \cdot \mathbf B_i,
  \label{eq:linear_site_energy}
\end{equation}
with a global parameter vector $\mathbf c$ and a feature vector $\mathbf B_i$ containing all ACE basis functions associated with atom $i$~\cite{Drautz2019,bochkarev2023acepotentials}.

To obtain a finite model we apply an a priori sparsification based on a total-degree criterion. Writing $k_t = (z_t,n_t,l_t,m_t)$, we define the mixed degree
\begin{equation}
  d(k_1,\dots,k_\nu)
    = \sum_{t=1}^{\nu} \left( n_t + w_L\, l_t \right),
  \label{eq:total_degree}
\end{equation}

and retain only those tuples with $d(k_1,\dots,k_\nu) \leq \texttt{totaldegree}(\nu)$. 
Here $w_L$ is a fixed weight controlling the relative contribution of angular and radial degrees in the total-degree truncation; throughout this work we use the default value $w_L = 1.5$. 
By contrast, $\texttt{totaldegree}(\nu)$ is varied and determines the overall basis size for each correlation order~$\nu$~\cite{bochkarev2023acepotentials}.

For a given configuration $R$, the linear contributions from all atomic environments are assembled into a design matrix $A_R$ that maps the parameter vector $\mathbf c$ to the stacked observables (energy, forces, stresses) for that structure. Concatenating all configurations in the training set $\mathcal R$ gives the global design matrix $A$ and an observation vector $\mathbf y$ such that
\begin{equation}
  \mathbf y \approx A\, \mathbf c,
  \label{eq:design_matrix_relation}
\end{equation}
which provides the starting point for the Bayesian fitting and leverage-score subsampling described below.

\subsection{Model fitting procedure}
\label{subsec:model_fitting}

The ACE parameters $\mathbf c$ are determined by fitting to a reference
dataset $\mathcal R$ of ASSYST-generated configurations with DFT energies and forces~\cite{Poul2025}. For each configuration $R \in \mathcal R$ we denote by $E_R$, $\mathbf F_R$, and $\mathbf V_R$ the reference total energy, forces, and virial stress, and by $E(\mathbf c;R)$, $\mathbf F(\mathbf c;R)$, and $\mathbf V(\mathbf c;R)$ the corresponding ACE predictions.

Following Ref.~\cite{bochkarev2023acepotentials}, we define a weighted least-squares objective
\begin{align}
  L(\mathbf c)
    &= \sum_{R \in \mathcal R}
       \Big[
         w_{E,R}^2\, \bigl| E(\mathbf c;R) - E_R \bigr|^2
\notag\\
    &\qquad\qquad
         +\, w_{F,R}^2\, \bigl\| \mathbf F(\mathbf c;R)
                               - \mathbf F_R \bigr\|^2
\notag\\
    &\qquad\qquad
         +\, w_{V,R}^2\, \bigl\| \mathbf V(\mathbf c;R)
                               - \mathbf V_R \bigr\|^2
       \Big],
  \label{eq:ls_objective}
\end{align}

where $w_{E,R}$, $w_{F,R}$, and $w_{V,R}$ control the relative importance of energies, forces, and stresses for each configuration $R$. 
In this work, we use constant weights $w_{E,R} = 1.0$ and $w_{F,R} = 0.1$ for all configurations, while stress data are not included ($w_{V,R} = 0$). 
Let $W$ denote the corresponding diagonal weight operator acting on the stacked observation vector $\mathbf y$, and let $A$ be the global design matrix constructed in Eq.~\eqref{eq:design_matrix_relation}. 
Then Eq.~\eqref{eq:ls_objective} can be written compactly as

\begin{equation}
  \mathbf c^\star
   = \arg\min_{\mathbf c}
     \bigl\| W (\mathbf y - A \mathbf c) \bigr\|^2 .
  \label{eq:weighted_ls}
\end{equation}

To regularize the fit and encode smoothness of the potential energy surface, we use a diagonal operator $\Gamma$ that assigns each basis function $k$ an effective degree $\gamma_k$ based on its radial and angular indices~\cite{bochkarev2023acepotentials}. This defines the quadratic penalty $\|\Gamma \mathbf c\|^2$ and leads to the Tikhonov-regularized problem
\begin{equation}
  \mathbf c^\star
    = \arg\min_{\mathbf c}
      \left(
        \bigl\| W (\mathbf y - A \mathbf c) \bigr\|^2
        + \lambda\, \|\Gamma \mathbf c\|^2
      \right),
  \label{eq:ridge_regression}
\end{equation}
with regularization strength $\lambda$.

We solve Eq.~\eqref{eq:ridge_regression} in the framework of Bayesian linear regression (BLR) as implemented in \textsc{ACEpotentials.jl}~\cite{bochkarev2023acepotentials}. In this interpretation, $\mathbf c$ is assigned a Gaussian prior
\begin{equation}
  \mathbf c \sim \mathcal N(\mathbf 0, \Sigma_0),
\end{equation}
with precision proportional to $\Gamma^\mathsf T \Gamma$ (smoothness prior), and the data likelihood is Gaussian with precision $\beta$ acting on $W(\mathbf y - A \mathbf c)$.

The posterior distribution of the parameters is again Gaussian,
\begin{equation}
  \mathbf c \,\big|\, A,\mathbf y
    \sim \mathcal N(\boldsymbol\mu, \Sigma),
\end{equation}
with covariance
\begin{equation}
  \Sigma
    = \Bigl(
        \beta\, A^\mathsf T W^2 A
        + \Sigma_0^{-1}
      \Bigr)^{-1},
  \label{eq:blr_covariance}
\end{equation}
and posterior mean
\begin{equation}
  \boldsymbol\mu
    = \beta\, \Sigma\, A^\mathsf T W^2 \mathbf y .
  \label{eq:blr_mean}
\end{equation}
In all numerical experiments reported here we use the posterior mean $\boldsymbol\mu$ as the deployed ACE model. The BLR formulation provides stable parameter estimates for the high-dimensional ACE basis, is robust in the low-data regime relevant for active learning, and naturally exposes parameter covariances that can be exploited in future uncertainty-quantification and model-selection strategies.

\subsection{Subset selection strategies}
\label{subsec:subset_selection}
\subsubsection{Configuration-level leverage scores}
\label{subsec:leverage_scores}

Within the linear ACE framework, fitting the model reduces to a weighted linear regression problem of the form
\[
\mathbf y \approx A \mathbf c,
\]
where each row of the global design matrix $A$ contains the ACE basis functions evaluated for a single observable (an energy, force component, or stress component), and $\mathbf c$ denotes the vector of ACE coefficients. Thus, the stacked vector $\mathbf y$ collects all observables from all configurations in the candidate pool.

The influence of individual observations on the fitted parameters can be quantified by statistical leverage scores~\cite{AL_Leverage_2018,OED_MLIP_2025}. 
Leverage-based sub-sampling has previously been employed to reduce training-set size and analyze cost–accuracy trade-offs in linear MLIPs~\cite{BAGHISHOV2025332}. 
Let $W$ denote the diagonal weight operator introduced in Eq.~\eqref{eq:weighted_ls}, and define the weighted design matrix
\begin{equation}
  \widetilde A = W A.
\end{equation}
In the unregularized least-squares setting, the classical leverage scores are the diagonal entries of the hat matrix
\begin{equation}
  H
  = \widetilde A
    \left( \widetilde A^{\mathsf T} \widetilde A \right)^{-1}
    \widetilde A^{\mathsf T},
\end{equation}
so that
\begin{equation}
  \ell_i = H_{ii},
\end{equation}

where $i$ indexes a \emph{row} of $\widetilde A$, i.e.\ a single energy, force, or stress component. Each row corresponds to a weighted ACE descriptor vector evaluated for one observable and can therefore be interpreted as a point in feature space. The hat matrix represents the projection onto the column space of $\widetilde A$, and the diagonal entry $\ell_i$ quantifies how strongly the $i$-th row contributes to spanning that space.

Rows with large $\ell_i$ introduce directions in descriptor space that are weakly represented by the remaining data and therefore exert strong influence on the fitted parameters. In the Bayesian linear setting, such rows also correspond to observations that most reduce the posterior covariance of the parameter vector. Since the differential entropy of a Gaussian posterior is proportional to $\log \det \Sigma$, selecting high-leverage rows can be interpreted as approximately maximizing information gain, or equivalently minimizing posterior entropy, in parameter space~\cite{OED_MLIP_2025,PCovR_CUR_2021}. 
In this context, ASSYST provides a descriptor-agnostic and transferable pool of configurations, while leverage-based selection refines this pool for a given ACE representation by identifying configurations that optimally span its descriptor space.

Our ACE fits, however, employ the Tikhonov-regularized Bayesian formulation of Eq.~\eqref{eq:ridge_regression}, with smoothness prior $\Gamma$ and prior precision $\Sigma_0^{-1}$, leading to the posterior covariance in Eq.~\eqref{eq:blr_covariance}. In this case it is natural to define \emph{regularized} leverage scores by replacing $\widetilde A^{\mathsf T}\widetilde A$ with the posterior precision,
\begin{equation}
  H_\lambda
    = \widetilde A
      \left(
        \beta\, \widetilde A^{\mathsf T} \widetilde A
        + \Sigma_0^{-1}
      \right)^{-1}
      \widetilde A^{\mathsf T},
\end{equation}
and we use the diagonal elements
\begin{equation}
  h_i = (H_\lambda)_{ii}
\end{equation}

as our working definition of leverage. Crucially, these scores depend only on the feature geometry ($\widetilde A$) and the regularization, and are therefore \emph{label-free}: they can be computed without access to the reference targets $\mathbf y$, i.e.\ before any DFT energies or forces have been evaluated~\cite{OED_MLIP_2025}.

Since our aim is to select \emph{configurations} rather than individual observables, we aggregate row-level leverage scores into a configuration-level quantity. For a configuration $R$, let $\mathcal I_R$ denote the set of row indices in $\widetilde A$ corresponding to its energy, all force components, and (where used) stress components. The configuration-level leverage is then defined as
\begin{equation}
  L_R = \sum_{i \in \mathcal I_R} h_i.
  \label{eq:config_leverage}
\end{equation}
Since ASSYST configurations consist of small periodic cells with comparable numbers of atoms, variations in $|\mathcal I_R|$ are limited and do not introduce a systematic bias in the ranking. In this setting, the sum in Eq.~\eqref{eq:config_leverage} provides a consistent measure of the total information content of each configuration.

Configurations with large $L_R$ probe under-represented directions in ACE feature space and are therefore expected to be particularly informative for the fit.

\subsubsection{Leverage-guided CUR and block-CUR subsampling} \label{subsec:cur_blockcur}

We now use these configuration-level leverage scores to construct compact, informative training subsets. Our strategy is inspired by CUR-type decompositions and optimal experimental design for high-dimensional feature matrices~\cite{PCovR_CUR_2021,BayesSelection_2025}:
training-set design is viewed as an \emph{offline} subset-selection problem on the ACE design matrix, where $L_R$ serves as a proxy for information content.

\paragraph*{Energy-CUR (``CUR'') mode.}

In the first mode, we operate on an energy-only feature matrix. For each configuration $R$ we retain a single row in $\widetilde A^{(E)}$ corresponding to its total energy; forces and stresses are omitted. This yields regularized leverage scores $h_i^{(E)}$ at the row level and configuration scores $L_R^{(E)}$ defined analogously to Eq.~\eqref{eq:config_leverage} but with $\mathcal I_R$ containing only the energy row. We then select a training subset $\mathcal R_{\mathrm{train}}$ from the candidate pool $\mathcal R_{\mathrm{pool}}$ by sampling without replacement with probabilities
\begin{equation}
  p_R^{(E)}
    = \frac{L_R^{(E)}}{\sum_{R' \in \mathcal R_{\mathrm{pool}}} L_{R'}^{(E)}},
\end{equation}
until a prescribed training fraction $f_{\mathrm{train}}$ of the pool has been selected.
This ``CUR'' mode mimics a row-selection step on an energy-only ACE design matrix and provides a simple, computationally cheap criterion that focuses on variation of total energies in descriptor space.

\paragraph*{Energy+force (``block-CUR'') mode.}

In the second mode, we retain all rows associated with each configuration, including energies and all Cartesian force components when forming $H_\lambda$. Configuration leverages $L_R$ are then given by Eq.~\eqref{eq:config_leverage} with $\mathcal I_R$ containing all observables for $R$. This ``block-CUR'' perspective treats each configuration as a block of correlated rows and favours those whose combined energy- force information most strongly constrains the ACE parameters~\cite{PCovR_CUR_2021,BayesSelection_2025}.

Selection proceeds as in the energy-CUR case, but with probabilities
\begin{equation}
  p_R^{(\mathrm{block})}
    = \frac{L_R}{\sum_{R' \in \mathcal R_{\mathrm{pool}}} L_{R'}},
\end{equation}
again sampling without replacement until the desired training fraction is reached. 

\subsubsection{Baseline selection strategies}
\label{Baseline_selection_strategies}

To benchmark the leverage-based, label-free methods introduced below, we consider three reference selection strategies.

All methods operate within the same incremental labelling protocol. For each run, the training set is initialized by randomly sampling 10\% of the available configurations from the ASSYST pool. A provisional ACE model is trained on this initial labelled subset.

Random selection then continues by sampling additional configurations uniformly from the remaining pool.

Energy-based and force-based selection proceed differently. After training the provisional ACE model on the current labelled subset, the model is evaluated on the remaining configurations in the pool (for which DFT labels are available in the benchmarking setup). Configurations are ranked according to their residual with respect to DFT reference data.

Energy-based selection ranks configurations by the absolute energy residual,
\begin{equation}
s_E(R) = \left| E_{\mathrm{ACE}}(R) - E_{\mathrm{DFT}}(R) \right|,
\end{equation}
while force-based selection ranks configurations using the root-mean-square (RMS) force residual,
\begin{equation}
s_F(R) =
\frac{\| F_{\mathrm{ACE}}(R) - F_{\mathrm{DFT}}(R) \|_2}
{\sqrt{3 N_R}},
\end{equation}
where $N_R$ is the number of atoms in configuration $R$.

At each iteration, the configurations with the largest residuals are added to the training set, and the model is retrained. Because these strategies require access to DFT labels for ranking, they represent label-driven upper-bound baselines rather than practical offline selection schemes.

\subsection{DFT reference calculations}
\label{subsec:dft_reference}

All validation energies and forces used to benchmark the ACE models were recomputed with density functional theory (DFT) to ensure full internal consistency with the training targets. Only the atomic geometries of the ASSYST test set from Ref.~\cite{Poul2025} were reused; no MTP predictions or energies from that study were used in the present analysis.

DFT calculations were performed with VASP using the projector--augmented wave (PAW) method and the PBE exchange--correlation functional. Standard VASP PAW datasets were employed for Al and Cu, with valence configurations Al(3s$^2$3p$^1$) and Cu(3d$^{10}$4s$^1$). A plane-wave cutoff of $E_\mathrm{cut}=550$~eV was used, converging total energies to better than $10^{-4}$~eV/atom and forces to $\sim 10^{-3}$~eV/\AA.

Brillouin-zone sampling used $\Gamma$-centered Monkhorst--Pack meshes selected to satisfy $\mathrm{KSPACING}=0.06$~\AA$^{-1}$ (e.g., $22\times22\times22$ for fcc Al and $18\times18\times18$ for fcc Cu at equilibrium volume). For larger supercells (defects, surfaces, grain boundaries, and Al--Cu alloy superstructures), the k-point density was reduced while maintaining $\mathrm{KSPACING}\le 0.10$~\AA$^{-1}$. Electronic occupations were treated with first-order Methfessel--Paxton smearing ($\sigma=0.20$~eV). Structural relaxations were converged to forces below $10^{-3}$~eV/\AA\ and total-energy changes below $10^{-7}$~eV; for very large cells, \texttt{LREAL=Auto} was used and verified to introduce only a constant energy shift that cancels in the reported energy differences.

All equilibrium properties (lattice parameters, equations of state, alloy formation energies, defect energetics, and surface/grain-boundary energies) were obtained from fully relaxed geometries followed by static single-point calculations. These DFT results provide the sole reference for assessing all linear ACE models reported here.

\section{Results}
\label{sec:results}
\subsection{Hyperparameter sensitivity of linear ACE models}
\label{subsec:res_hyperparam_sensitivity}

Before analysing training-set design, we first assess the sensitivity of linear ACE model accuracy to key hyperparameters controlling basis size and interaction range. We focus on three primary ACE hyperparameters: the body order $\nu$, the total-degree truncation parameter \texttt{totaldegree}, and the radial cutoff $r_{\mathrm{cut}}$. For this study, the full ASSYST dataset was randomly partitioned into 80\% training data and 20\% validation data. The same fixed split was used for all hyperparameter combinations to ensure a consistent comparison.  This analysis serves as a baseline to ensure that subsequent differences between subset-selection strategies do not arise from hyperparameter tuning.

For each hyperparameter choice, a linear ACE model was trained on the 80\% training subset using identical fitting protocols and evaluated on the held-out 20\% validation set in terms of energy and force root-mean-square errors (RMSE). To focus on practically relevant accuracy regimes, we restrict attention to models achieving energy RMSEs below 50 meV/atom on the validation set.

\begin{figure*}[t]
  \centering
  \includegraphics[scale=0.50]{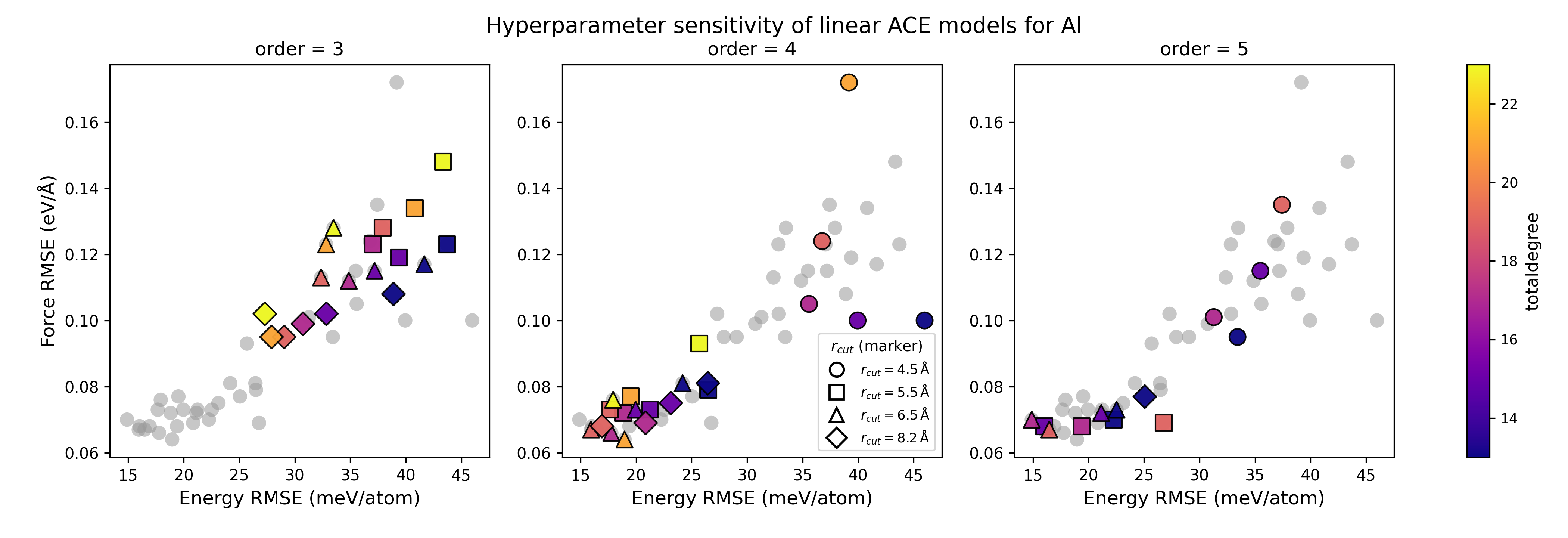}\par\vspace{0.6em}
  \includegraphics[scale=0.50]{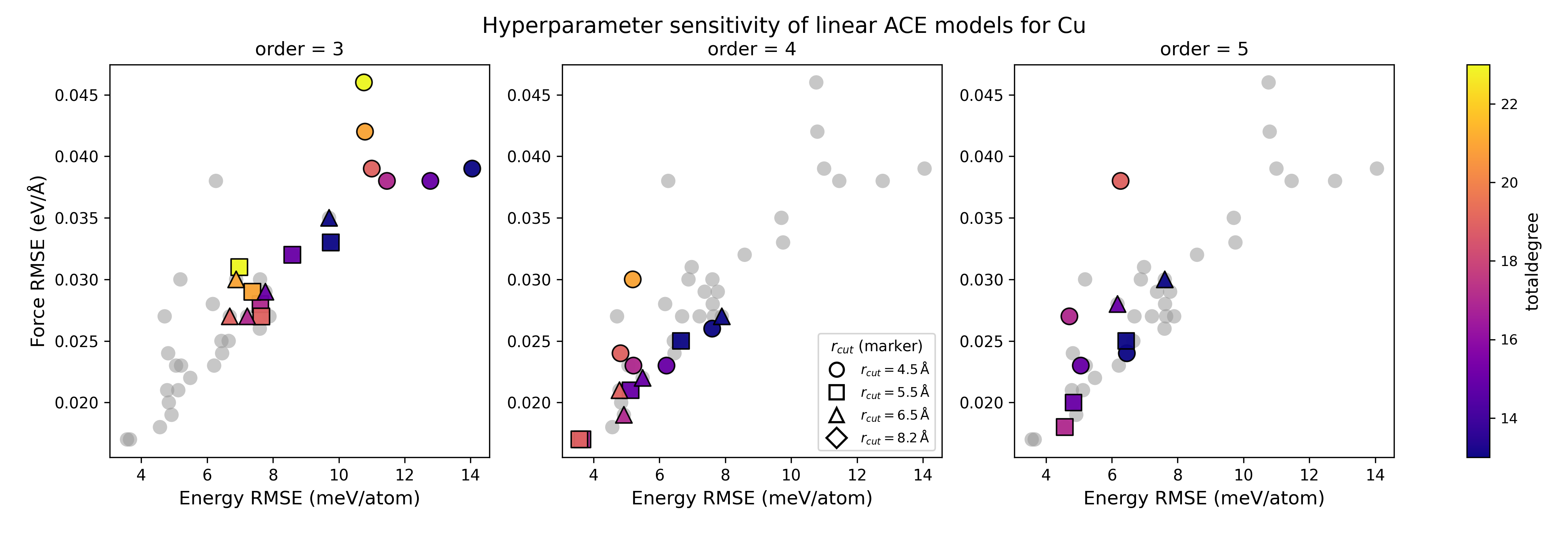}
    \caption{Hyperparameter sensitivity of linear ACE models for Al (top) and Cu (bottom), evaluated using held-out test RMSEs. For each body order ($\nu = 3,4,5$), colored markers indicate models with that order, with color encoding \texttt{totaldegree} and marker size reflecting the cutoff radius $r_{\mathrm{cut}}$. Gray markers correspond to models from other body orders and are shown for reference. Because the errors are evaluated on unseen test configurations, the dependence on \texttt{totaldegree} is not necessarily monotonic; high-complexity models can show larger test errors due to overfitting.}
  \label{fig:hyperparam_sensitivity_Al_Cu}
\end{figure*}

Figure~\ref{fig:hyperparam_sensitivity_Al_Cu} summarizes the resulting error landscapes for Al and Cu in the $(E_{\mathrm{RMSE}},F_{\mathrm{RMSE}})$ plane. Each panel corresponds to a fixed body order $\nu$. Colored markers denote models with that body order, with color encoding \texttt{totaldegree} and marker size reflecting $r_{\mathrm{cut}}$. Gray markers show hyperparameter combinations from the other body orders to provide global context for the attainable error range. Several consistent trends emerge across both materials:
\begin{itemize}
    \item Increasing $r_{\mathrm{cut}}$ systematically reduces both energy and force RMSEs, highlighting the importance of capturing longer-range interactions even within a linear ACE representation.
    \item Increasing \texttt{totaldegree} generally lowers the test errors up to a saturation regime. At the highest basis sizes the trend is not strictly monotonic, which is consistent with overfitting: additional basis functions may improve the fit to the training data without improving generalization to the held-out test set.
    \item Higher body order primarily benefits force accuracy at fixed energy error, indicating that many-body correlations play a larger role in reproducing force information than total energies.
\end{itemize}

The spread of points also reveals diminishing returns beyond $\nu = 4$ and shows that increasing model complexity does not always improve test performance, particularly when the larger basis begins to overfit the training data. 

Notably, the Cu models lie on a lower-error manifold than the Al models for comparable hyperparameters. This trend is consistent with previous studies reporting differences in model complexity across materials, although a detailed analysis of the underlying physical origin is beyond the scope of the present work.

Importantly, this analysis reveals a broad hyperparameter region in which comparable accuracy can be achieved. This ensures that the performance differences observed between random sampling, energy-based selection, and leverage-guided strategies do not arise from fine-tuning of ACE hyperparameters, but instead reflect genuine differences in the information content of the selected configurations. We therefore fix a single set of hyperparameters in the following for clarity of comparison. For all subset-selection comparisons reported below, we use body order $\nu=3$, total-degree truncation $\texttt{td}=17$, and cutoff radius $r_{\mathrm{cut}}=6.5$~\AA. This fixed architecture was chosen from the low-error region of the hyperparameter scan and is used consistently across the subset-selection comparisons.


\subsection{Learning efficiency for Al: energy and force test errors}
\label{subsec:res_al_efficiency}

We next quantify the data efficiency of different subset-selection strategies for elemental Al under a controlled iterative data-acquisition protocol designed to mimic progressive DFT labeling. The ASSYST pool contains approximately $N_{\mathrm{tot}}\approx 9\times 10^3$ configurations. In each run, $20\%$ of the data are reserved as an independent test set and never used during fitting. From the remaining configurations, the training set is initialized with $10\%$ of the available data and iteratively expanded by adding $10\%$ of the remaining pool according to the chosen selection rule (random, energy-based, force-based, leverage-CUR, or leverage block-CUR). This protocol enables a direct comparison of how efficiently each strategy improves predictive accuracy as additional DFT labels are acquired.

To compare methods at equal data budgets, we report learning curves as a function of the labelled fraction
\begin{equation}
f_{\mathrm{lab}}
= \frac{N_{\mathrm{train}}}{N_{\mathrm{train}} + N_{\mathrm{pool}}},
\end{equation}
plotted as $100\,f_{\mathrm{lab}}$. To account for statistical variability arising from random 80/20 train–test splits, random initialization of the initial 10\% training subset, and stochastic elements in the selection rules, each experiment is repeated five times with independent random seeds. For each method, we report RMSE values for energies and forces, with shaded regions indicating 95\% confidence intervals computed across the five runs. Test RMSE measures generalization to unseen configurations and therefore reflects the physical transferability of the fitted potential.

\begin{figure}[t]
  \centering
  \includegraphics[scale=0.52]{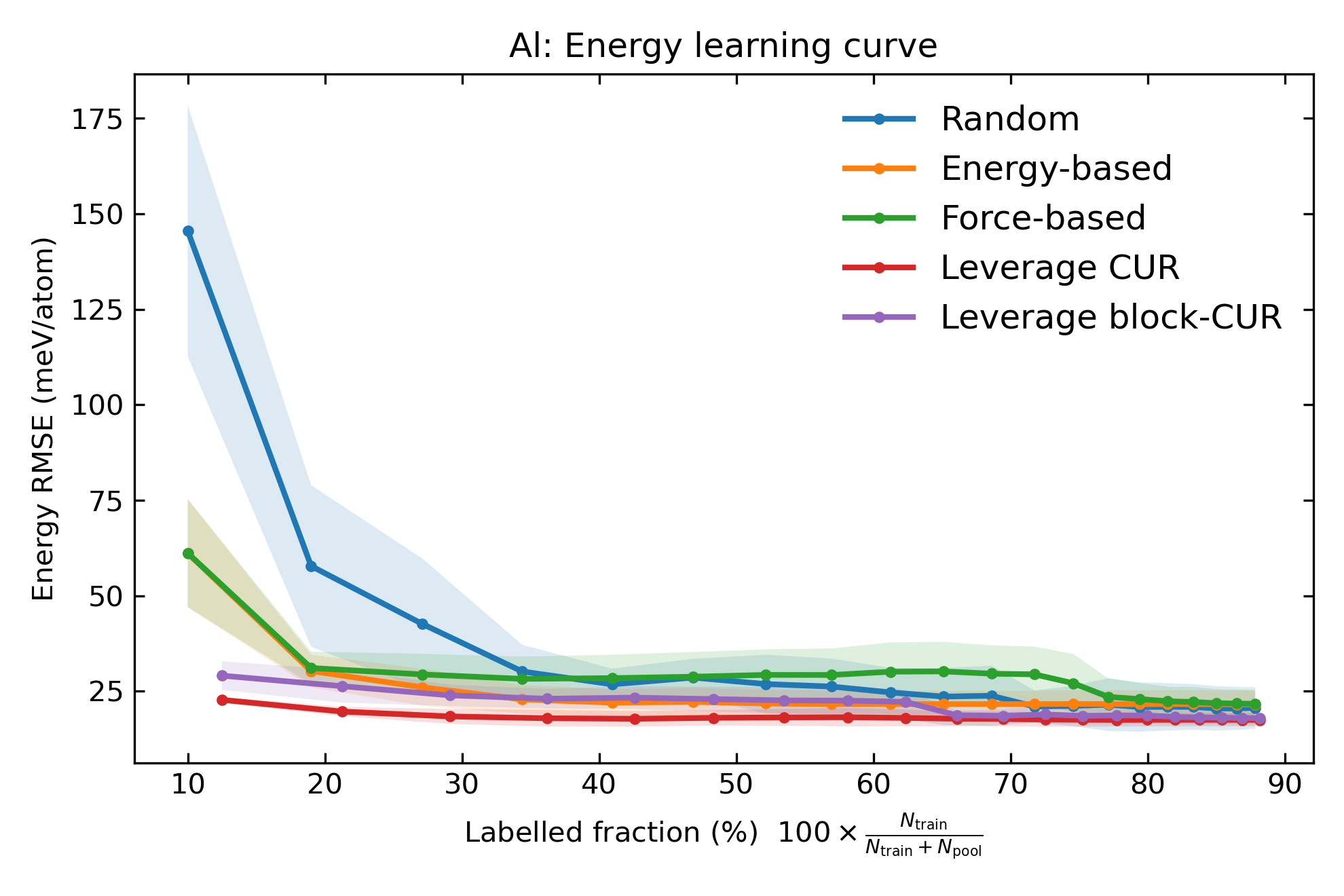}\par\vspace{0.6em}
  \includegraphics[scale=0.52]{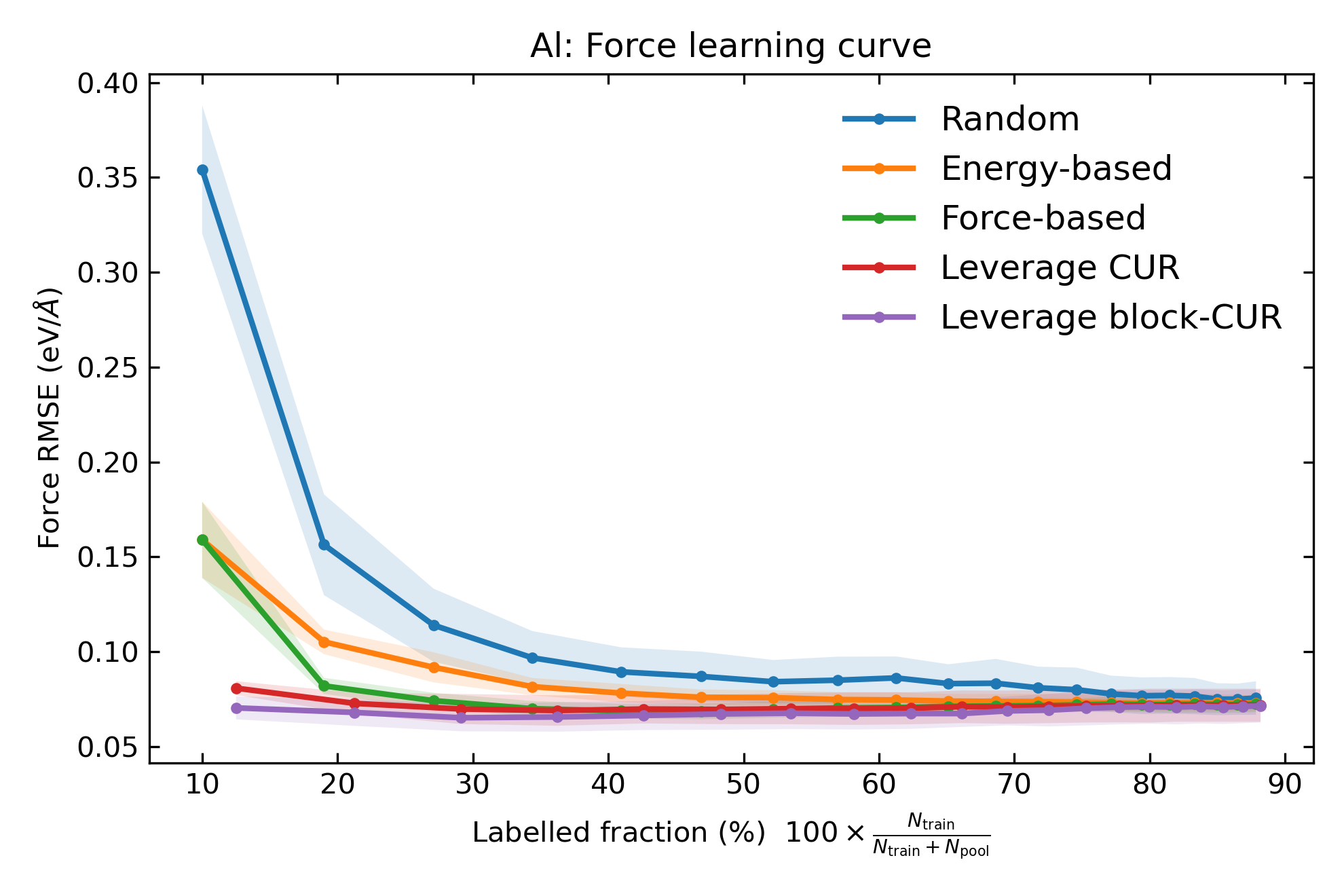}
   \caption{Test energy and force learning curves for Al. Solid lines denote the mean RMSE over five independent runs (different random train–test splits and initial training subsets), and shaded regions indicate $\pm 95\%$ confidence intervals. Results are plotted versus labelled fraction $100\,f_{\mathrm{lab}}$.}
  \label{fig:al_learning_curves_Al}
\end{figure}

Figure~\ref{fig:al_learning_curves_Al} shows the resulting learning curves. For all methods, both energy and force errors decrease monotonically with increasing labelled fraction, confirming that additional DFT data improve the fitted ACE model. However, convergence rates differ markedly between selection strategies. Random sampling provides a baseline corresponding to uniform exploration of the configuration pool. Energy-based selection accelerates early reduction of energy errors but yields limited improvement in force accuracy. Force-based selection performs competitively at small labelled fractions but exhibits larger run-to-run variability and saturates at higher errors, suggesting that selecting configurations solely by large force magnitudes does not efficiently span descriptor space.

In contrast, leverage-based strategies provide improved data efficiency. Both CUR and block-CUR consistently achieve lower test errors than heuristic baselines across most labelled fractions, with the advantage most pronounced for $f_{\mathrm{lab}}\lesssim 40\%$, the regime most relevant for reducing DFT cost. Energy RMSEs for CUR and block-CUR are nearly indistinguishable, indicating that both strategies effectively constrain the dominant energy modes of the ACE parameter space. A clear distinction emerges for forces: block-CUR consistently yields the lowest force errors and reaches its accuracy plateau at $f_{\mathrm{lab}}\approx 30\text{--}40\%$, whereas random sampling requires a substantially larger labelled fraction to reach comparable performance.

The distinction becomes clear when considering what each observable constrains: energies constrain global features of the potential-energy surface, whereas forces probe local curvature and therefore impose significantly stronger constraints on model parameters. By incorporating the coupled energy--force block structure of each configuration, block-CUR preferentially selects configurations that constrain both total-energy variations and local force responses, contributing to faster convergence of force predictions.

Taken together, these results demonstrate that leverage-guided subset selection provides a more balanced and physically informative sampling of configuration space than scalar heuristics based solely on energy or force magnitudes. Energy-only CUR provides substantial gains in the low-labelled regime at low computational cost, while block-CUR further enhances robustness and transferability for applications requiring accurate forces, such as molecular dynamics, lattice dynamics, and defect migration. Overall, leverage-based selection achieves comparable predictive accuracy using substantially fewer DFT-labelled configurations, directly reducing the computational cost of MLIP development.

\subsection{Transferability to Cu and Al--Cu}
\label{subsec:res_transfer_cu_alcu}

To assess whether the subset-selection behaviour observed for Al extends to other systems, we applied the same training protocol and model architecture to Cu and Al--Cu systems. In each case, ACE models were trained on ASSYST-generated structure pools specific to the corresponding material, using identical hyperparameter settings and selection strategies. For Cu, we adopted the same linear ACE architecture as in the Al study (body order $\nu = 3$, total-degree truncation $\texttt{td}=17$, and cutoff $r_{\mathrm{cut}} = 6.5\,\text{\AA}$) to isolate the effect of dataset selection. A pool of 10{,}631 Cu structures generated by \textsc{Assyst} served as the candidate pool for iterative subset selection. While the architectural and selection framework was transferred from the Al benchmark, all ACE parameters were fitted exclusively using Cu reference data.

\begin{figure}[t]
  \centering
  \includegraphics[scale=0.52]{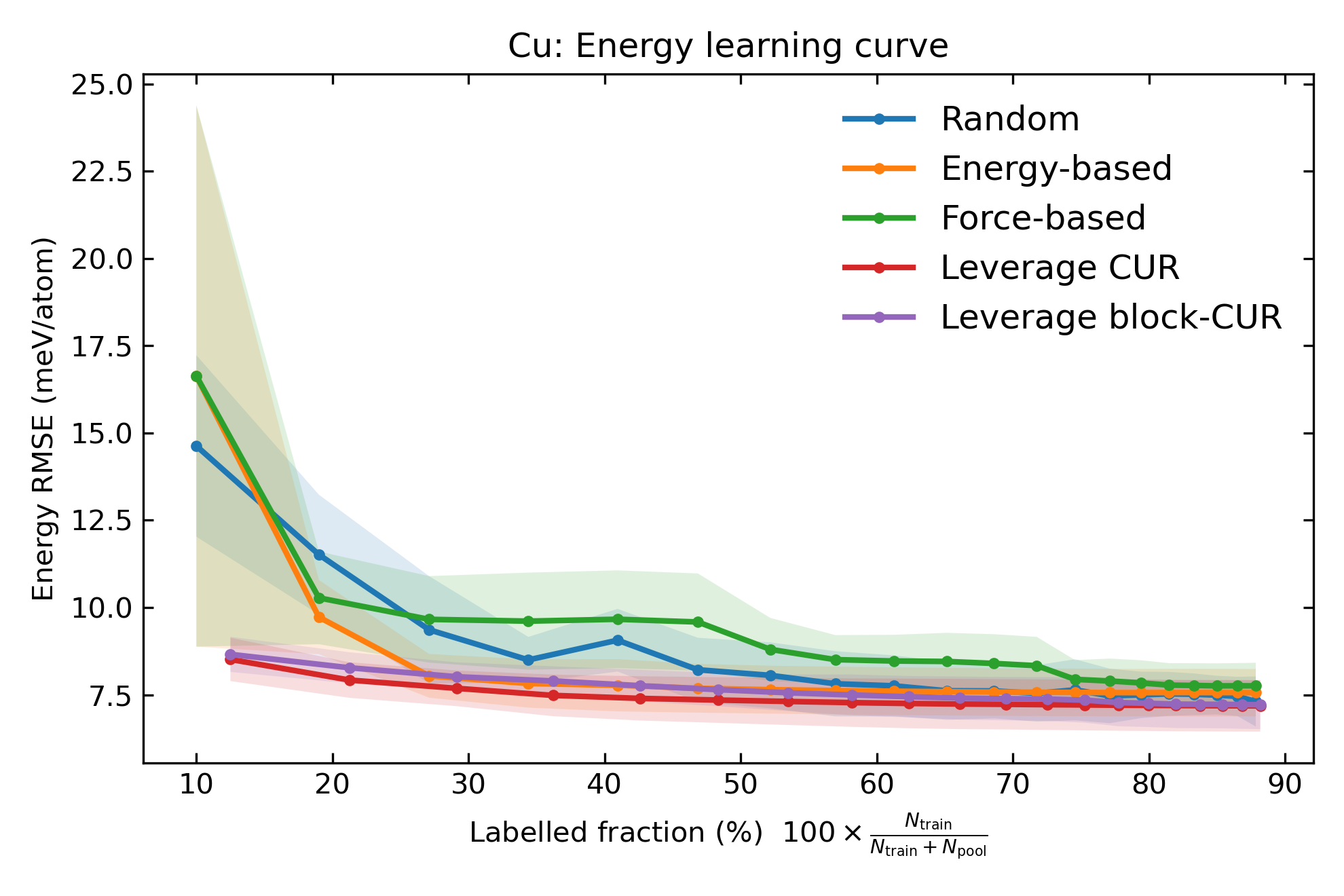}\par\vspace{0.6em}
  \includegraphics[scale=0.52]{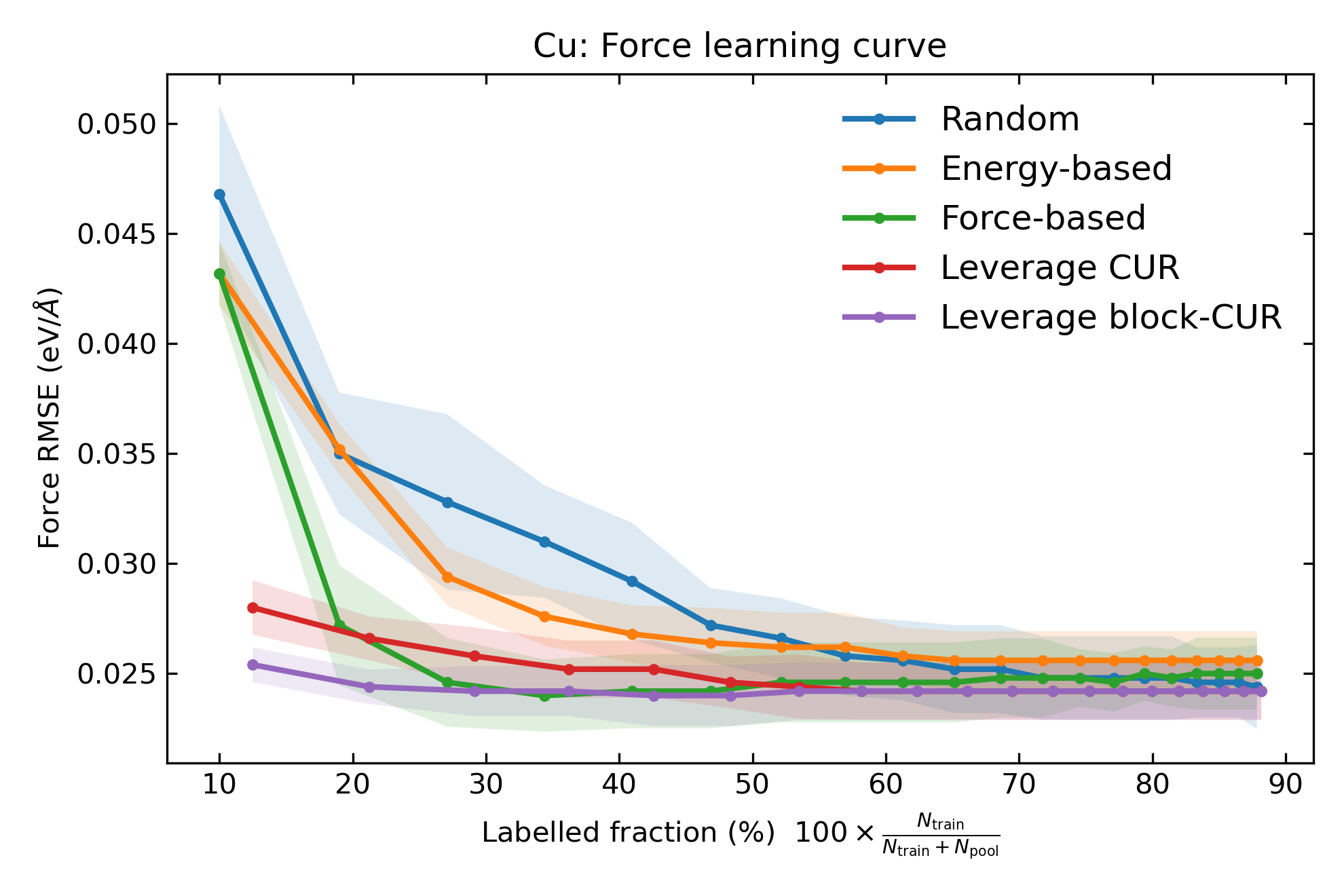}
  \caption{Energy and force learning curves on Cu, obtained from five independent runs of the iterative acquisition protocol. Solid lines correspond to the mean RMSE and the shaded regions denote $\pm 95\%$ confidence intervals. The labelled fraction is reported as $100\,f_{\mathrm{lab}}$.}
  \label{fig:al_learning_curves_Cu}
\end{figure}

Figure~\ref{fig:al_learning_curves_Cu} shows the learning curves for energy and force RMSE as a function of the labelled fraction. Across all fractions, both CUR-type strategies (CUR and block-CUR) exhibit faster convergence and reach the saturation regime at substantially smaller labelled fractions compared to the baseline methods. In particular, force RMSEs of $\sim 0.024\text{--}0.025\,\mathrm{eV/\text{\AA}}$ and energy RMSEs of $\sim 7\text{--}7.5\,\mathrm{meV/atom}$ are achieved once roughly $40\text{--}50\%$ of the pool has been labelled.

At higher labelled fractions, all methods approach similar error levels, as expected since the training sets become increasingly similar. Differences in this regime are therefore less significant. In contrast, at lower and intermediate labelled fractions, leverage-guided selection generally achieves lower errors and smoother convergence, indicating improved data efficiency.

Force-based selection reduces the force RMSE more rapidly at small labelled fractions ($10\text{--}20\%$) but plateaus at higher energy errors ($\approx 9\,\mathrm{meV/atom}$), indicating limited improvement in capturing bonded energetics. Random and energy-based selection exhibit slower convergence overall and require larger labelled fractions to reach comparable accuracy.

Importantly, each learning curve reports the mean over five independent runs of the iterative acquisition protocol, while the shaded bands denote $\pm 95\%$ confidence intervals, reflecting the robustness and reproducibility of each selection strategy (Fig.~\ref{fig:al_learning_curves_Cu}). The narrow confidence intervals associated with CUR-type methods highlight their stability and consistent performance, whereas the greater variability observed for force-based and random selection indicates reduced reliability in the transferred setting.

Taken together, these results indicate that CUR-type leverage sampling improves data efficiency in the Cu system by reaching comparable accuracy at smaller labeled fractions.

\begin{figure}[t]
  \centering
  \includegraphics[scale=0.52]{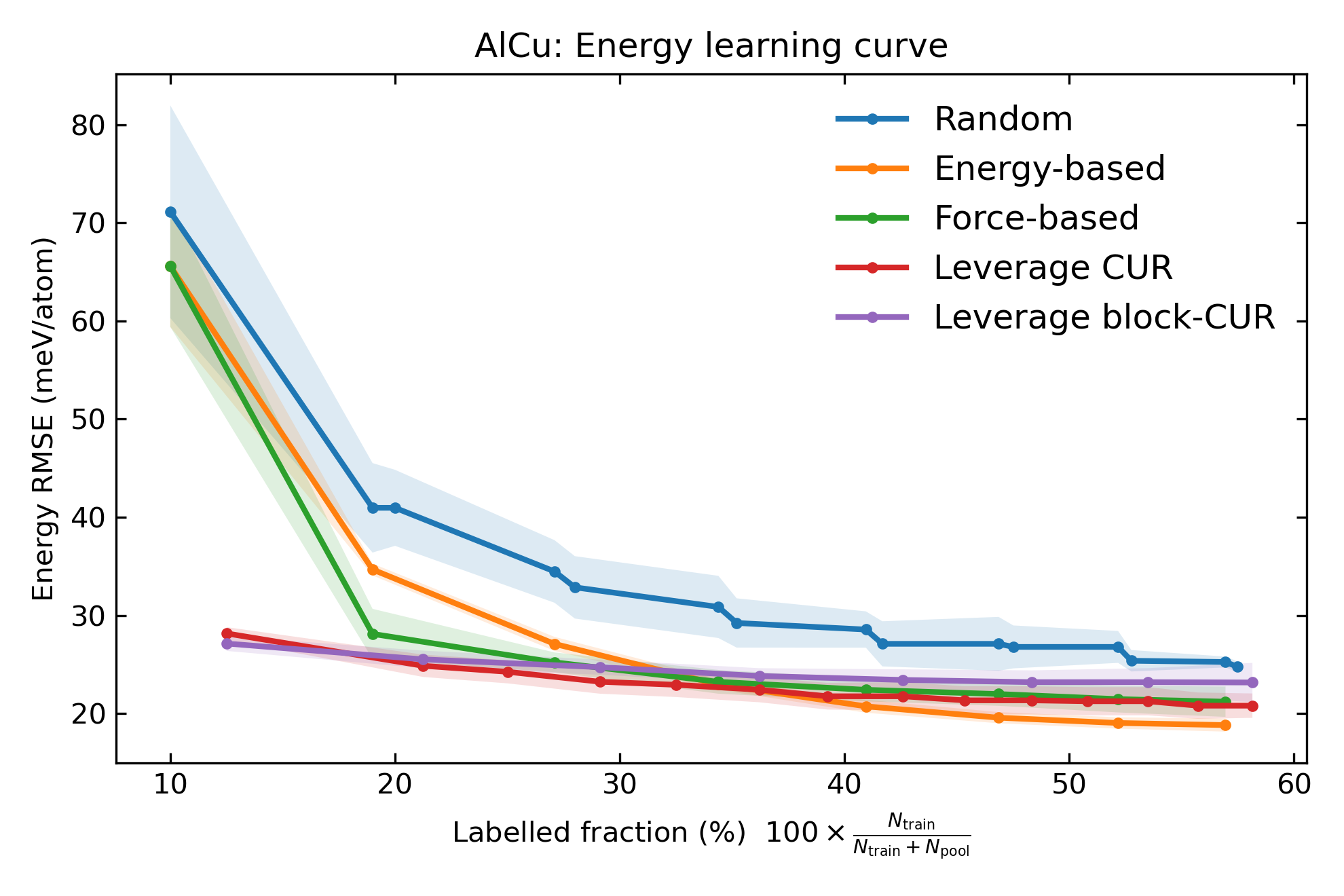}\par\vspace{0.6em}
  \includegraphics[scale=0.53]{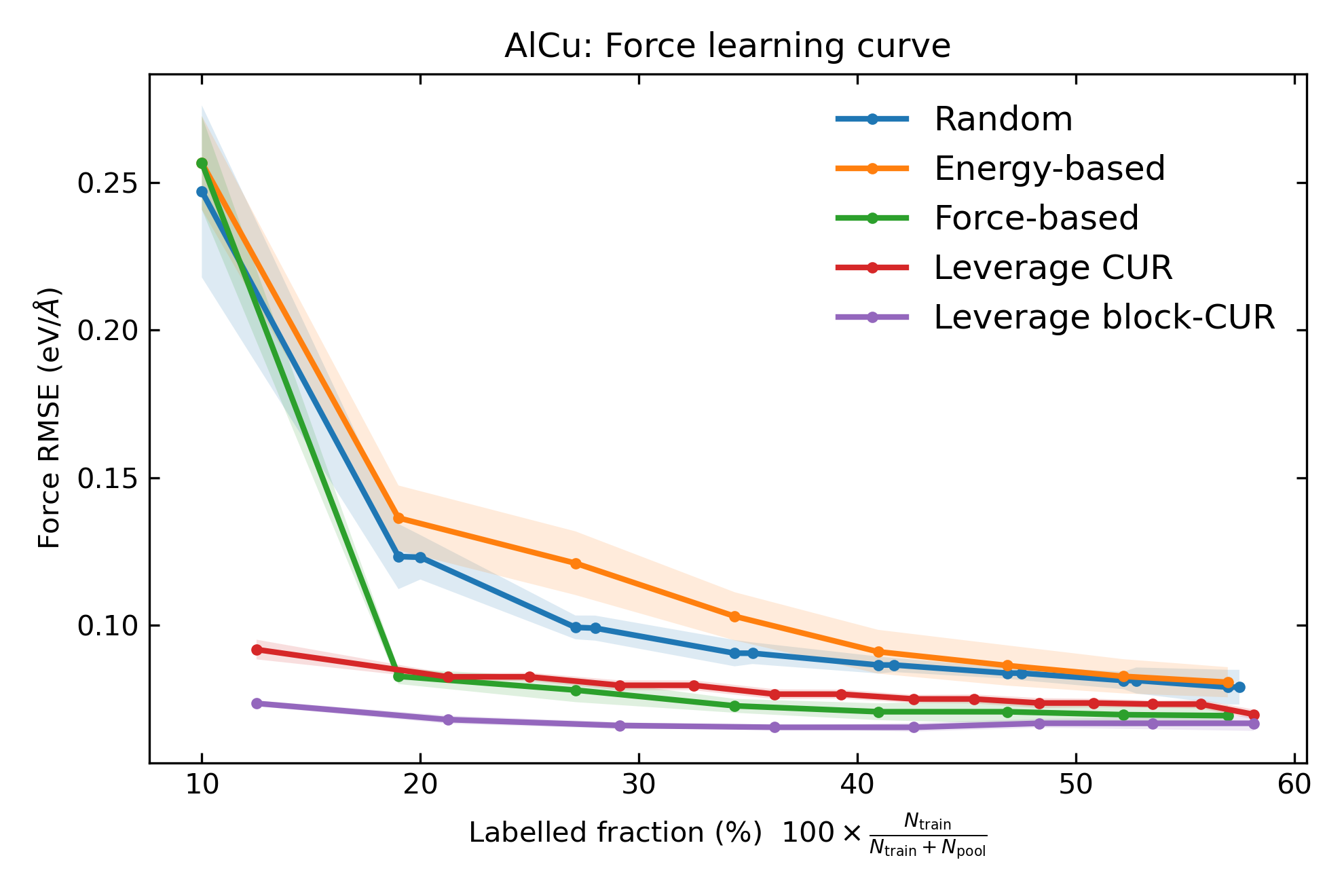}
  \caption{Test-set learning curves for Al--Cu alloys comparing five selection strategies. Solid lines denote the mean RMSE over five active-learning runs; shaded regions indicate 95\% confidence intervals. Labeled fraction is shown as $100\,f_{\mathrm{lab}}$.}
  \label{fig:al_learning_curves_AlCu}
\end{figure}

For Al--Cu alloys, we employed the same ACE architecture optimized for elemental Al (body order $\nu = 3$, total degree $\texttt{td}=17$, and cutoff $r_{\mathrm{cut}} = 6.5\,\text{\AA}$) and evaluated the five selection strategies on a pool of $21{,}330$ distinct Al--Cu configurations. Figure~\ref{fig:al_learning_curves_AlCu} shows the resulting test-set learning curves for energy and force RMSE as a function of the labelled fraction $100\,f_{\mathrm{lab}}$. Solid lines denote the mean over five independent runs, and shaded regions represent the corresponding $95\%$ confidence intervals.

For energy errors, all strategies exhibit monotonic improvement as the labelled fraction increases, indicating that additional DFT data refine the global energetics of the alloy potential-energy surface. At small labelled fractions, the CUR and block-CUR strategies provide a clear advantage, achieving lower energy RMSE than the heuristic baselines. This is consistent with improved coverage of structurally and chemically diverse environments early in the acquisition process. As the labelled fraction grows, the purely energy-based heuristic becomes competitive and eventually slightly surpasses the CUR-type methods, while force-based selection lies between these extremes. Random sampling shows the slowest convergence and consistently higher energy errors across the entire range.

The force learning curves show a clearer separation between methods.
Across all labeled fractions, CUR-type strategies achieve the lowest force RMSE, with block-CUR converging most rapidly and reaching the smallest asymptotic error, followed closely by CUR. Both methods remain below the energy-based, force-based, and random heuristics across most labelled fractions. This indicates that leverage-guided selection more effectively constrains the local curvature of the potential-energy surface, which is directly probed by force information and is particularly sensitive to chemically heterogeneous environments in the alloy.

Overall, the alloy results follow the qualitative trends observed in the elemental benchmarks. Leverage-based sampling reaches comparable force accuracy at smaller labelled fractions, while energy-based selection becomes competitive for total-energy accuracy once a sufficiently large fraction of the pool has been labelled. The narrow confidence intervals indicate that these trends are consistent across independent runs.

\subsection{Validation against independent DFT calculations}
\label{subsec:res_dft_validation}
\begin{figure*}[!htbp]
  \centering
  \includegraphics[scale=0.52]{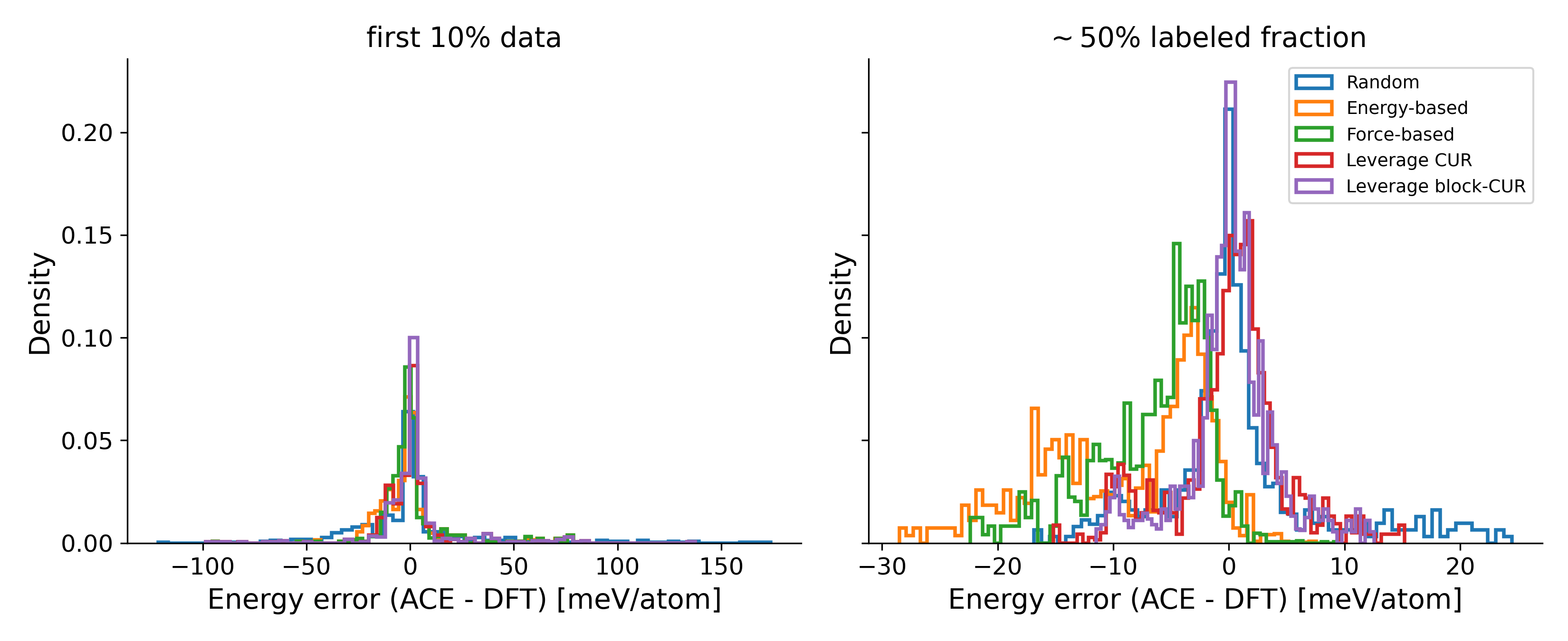}\par\vspace{0.8em}
  \includegraphics[scale=0.52]{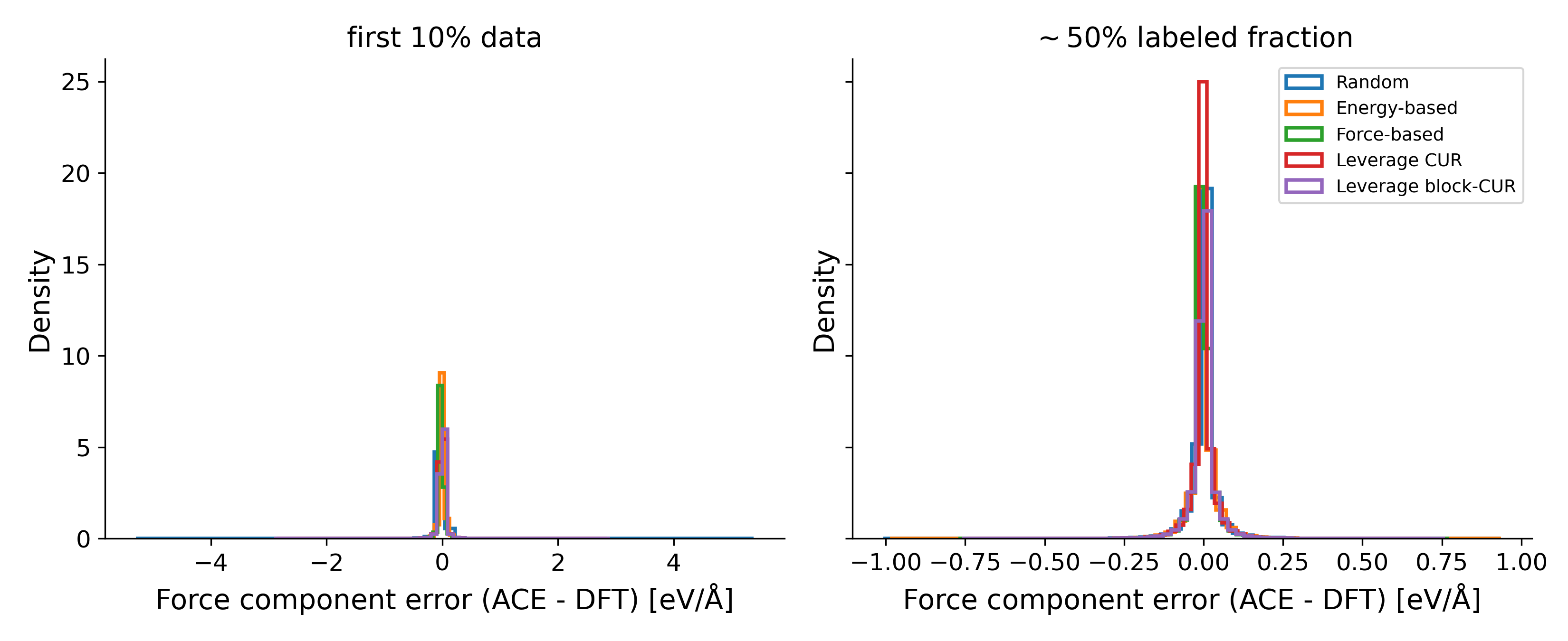}

  \caption{ Error distributions on the independent DFT validation set for the different subset-selection strategies.  The top panel shows per-atom energy errors and the bottom panel shows force-component errors, both defined as ACE $-$ DFT. Each plot compares the early low-data regime, corresponding to the first 10\% labelled data, with the approximately 50\% labelled regime. The error distributions narrow and become increasingly centered near zero as the labelled fraction increases, indicating improved agreement with the independent DFT reference data.}
  \label{fig:validation_error_distributions}
\end{figure*}

\begin{figure}[!htbp]
  \centering
  \includegraphics[scale=0.36]{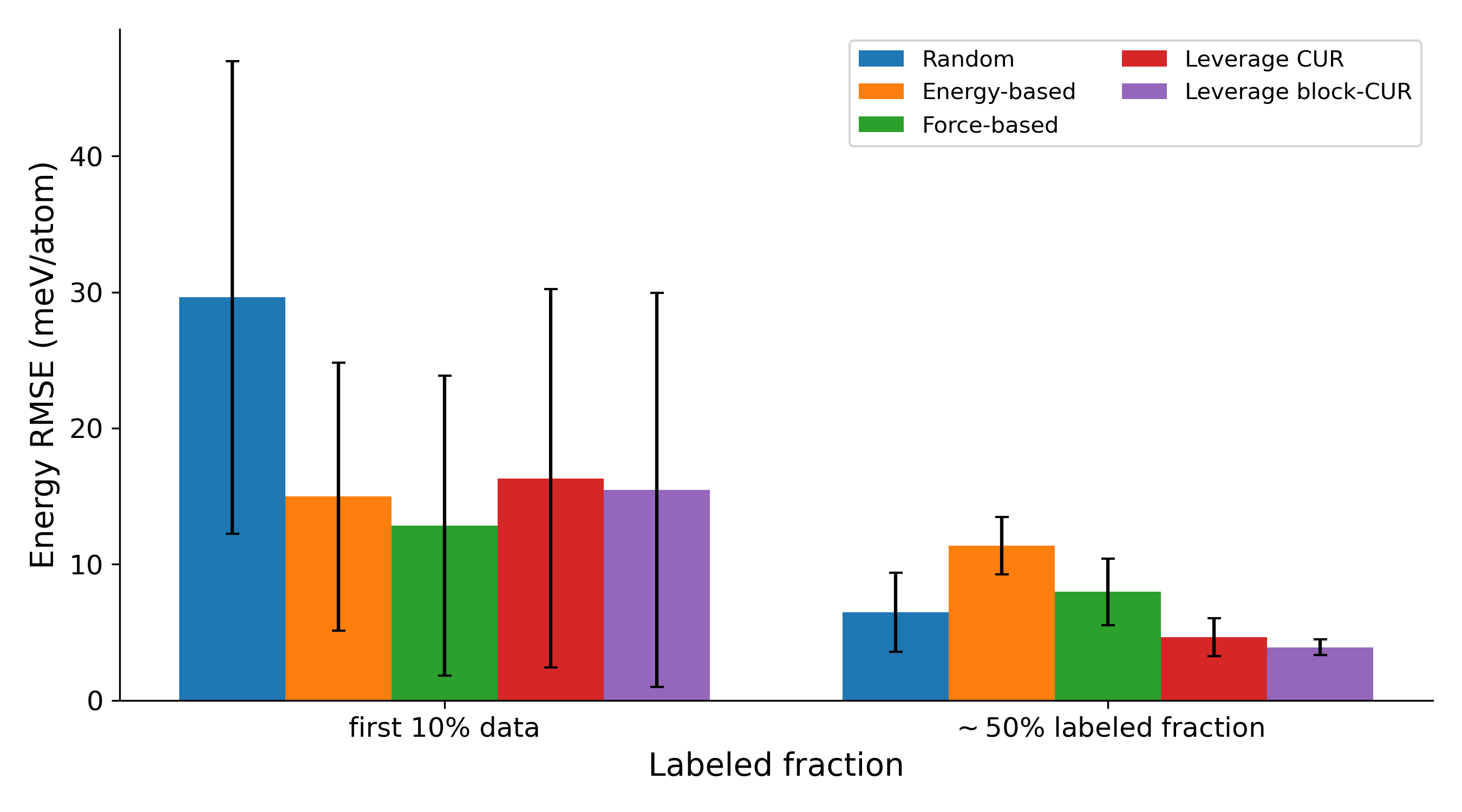}\par\vspace{0.8em}
  \includegraphics[scale=0.36]{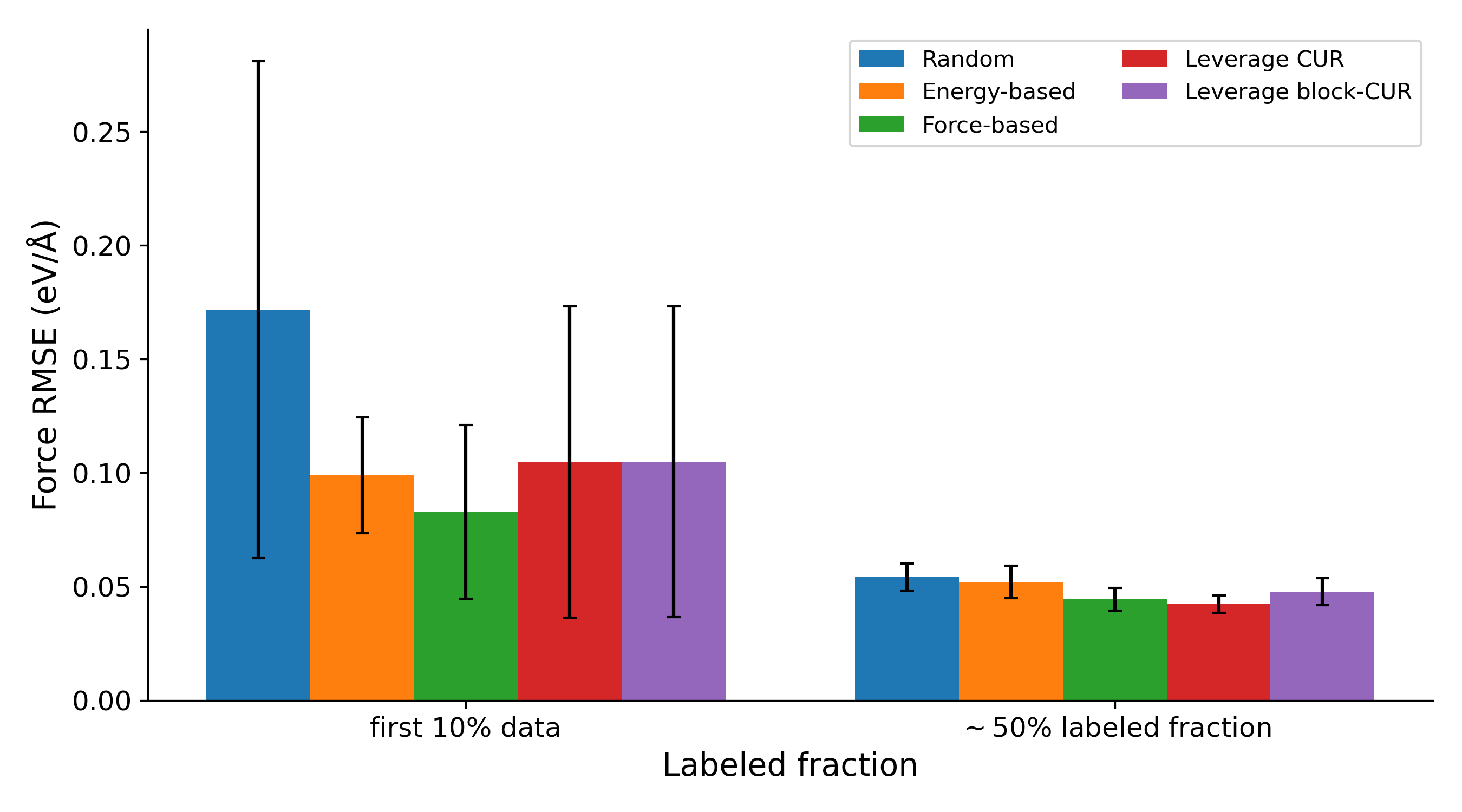}

  \caption{ RMSE summary on the independent DFT validation set for the different subset-selection strategies. The top panel reports the energy RMSE and the bottom panel reports the force RMSE for the first 10\% labelled data and for the approximately 50\% labelled regime.  Error bars indicate variation across the selected iteration windows.  The comparison shows that leverage-based CUR and block-CUR selection achieve competitive or lower RMSE values in the approximately 50\% labelled regime while using substantially reduced training-set sizes.}
  \label{fig:validation_rmse_bars}
\end{figure}

For Al, we validated all ACE models against an \emph{independent} DFT test set of 360 configurations comprising bulk crystals, symmetric tilt grain boundaries, free surfaces, point defects, and vacancy structures. 
To avoid the visual overlap that arises when several selection strategies are shown in parity plots, we summarize the validation results using both ACE--DFT error distributions and RMSE values. 
Figure~\ref{fig:validation_error_distributions} shows the distributions of per-atom energy errors and force-component errors in the early low-data regime and in the approximately 50\% labelled regime. 
Figure~\ref{fig:validation_rmse_bars} reports the corresponding energy and force RMSE values for the same comparison.

At the first 10\% labelled fraction, the validation errors are broad, particularly for random selection, reflecting the under-constrained nature of the low-data regime. 
As the labelled fraction increases to approximately 50\%, the error distributions narrow substantially and become more strongly centered near zero for both energies and forces. 
This behaviour confirms that the selected configurations progressively improve agreement with the independent DFT reference data rather than only reducing errors on the ASSYST train--test splits.

The RMSE summaries in Fig.~\ref{fig:validation_rmse_bars} show the same trend quantitatively. 
Energy and force RMSEs decrease strongly between the first 10\% labelled fraction and the approximately 50\% labelled regime. 
In the converged regime, leverage-based CUR and block-CUR achieve competitive or lower energy RMSEs than the heuristic baselines, while the force RMSEs remain comparable across methods. 
These results are consistent with the learning curves and indicate that leverage-guided subsets preserve predictive accuracy on independent bulk, surface, grain-boundary, and vacancy-containing structures.

Because the validation energies are reported on a per-atom basis, defect contributions can be partially averaged out in the global error distributions. 
For this reason, defect-specific validation metrics, including vacancy formation and grain-boundary energies, are analysed separately below.




For Cu, we examined whether the different selection strategies preserve the defect-relevant regions of configurational space. In addition to low bulk RMSE on ASSYST-generated structures, the models were evaluated against two sensitive defect observables: the vacancy formation energy in bulk Cu and the grain-boundary energy of a $\Sigma5(310)[001]$ symmetric tilt boundary. These quantities probe broken-symmetry environments that are sparsely represented in the training pool and therefore provide a stringent test of transferability.

As shown in Table~\ref{tab:cu_defect_surface_summary}, leverage-based selection yields markedly lower defect-energy errors. The CUR and block-CUR models achieve vacancy formation energy errors of $0.041 \pm 0.015$\,eV and $0.047 \pm 0.015$\,eV, respectively, whereas random, energy-based, and force-based strategies overestimate the vacancy energy by approximately $0.25$--$0.26$\,eV. For the $\Sigma5$ grain boundary, leverage-based models reduce the error to $0.0046$--$0.0051$\,eV/\AA$^2$, compared with $0.0067$--$0.0073$\,eV/\AA$^2$ for the heuristic baselines.

Importantly, these differences arise despite similar global energy and force RMSE across methods. Vacancy formation and grain-boundary energetics probe locally under-coordinated and symmetry-broken environments that are sparsely represented in the training pool. The substantially lower defect errors obtained with leverage-guided selection therefore indicate that the high-leverage subsets more effectively span these low-coordination regions of descriptor space, constraining the corresponding directions in parameter space beyond what is captured by aggregate bulk RMSE metrics alone.

\begin{table*}[t]
\centering
\small
\caption{Summary of defect and surface errors for Cu with respect to DFT.
Vacancy error is averaged over 3$\times$3$\times$3 and 4$\times$4$\times$4 cells.
Surface error is averaged over all surface orientations.
Global energy and forces values are taken as RMSE over all structures after ACE relaxation.
Values represent the mean over five independent runs, digits in the parentheses the standard deviation in the trailing digits.}
\label{tab:cu_defect_surface_summary}
\begin{tabular}{lcccccc}
\toprule
Method
& ${E_f^{vac}}$
& $\gamma_{\mathrm{GB}}$
& $\langle \gamma_{\mathrm{surf}} \rangle$
& Global $E$ 
& Global $F$  \\
 & eV & eV/\AA$^2$ & eV/\AA$^2$  & m\,eV/atom & m\,eV/\AA \\
\midrule

Random
& $0.246(6)$
& $0.0067(6)$
& $0.010(7)$
& $5$
& $0.5$ \\

Energy-based
& $0.260(6)$
& $0.0072(4)$
& $0.010(6)$
& $5$
& $0.5$ \\

Force-based
& $0.259(7)$
& $0.0073(4)$
& $0.009(6)$
& $4$
& $0.5$ \\

Leverage CUR
& $0.041(15)$
& $0.0046(6)$
& $0.006(7)$
& $3$
& $0.5$ \\

Leverage block-CUR
& $0.047(15)$
& $0.0051(5)$
& $0.007(7)$
& $3$
& $0.5$ \\

\bottomrule
\end{tabular}
\end{table*}

Table~\ref{tab:alcu_ordered_aggregate} reports aggregate defect errors for ordered Al--Cu intermetallic compounds relative to DFT. 
The complete Al–Cu training pool contains 21{,}330 configurations; the 50\% baselines therefore correspond to approximately 10{,}700 labeled structures, whereas the leverage-based models at 25\% are trained on only about 5{,}300 configurations.

Across the 50\% baselines, antisite formation energies exhibit large RMSEs in the range $0.22$–$0.28$\,eV, confirming that chemically ordered environments provide a stringent transferability test for linear ACE models. Vacancy formation energies show smaller but still non-negligible deviations ($0.15$–$0.18$\,eV), reflecting the sensitivity of vacancy energetics to sublattice specificity and local chemical asymmetry.

Reducing the labeled fraction to 25\% does not lead to a clear degradation in ordered-defect accuracy when leverage-guided selection is employed. Both CUR and block-CUR yield antisite and vacancy RMSEs that remain within the same overall range as the 50\% baseline models, despite a twofold reduction in training size. This suggests that the dominant chemically distinct environments in ordered compounds are already reasonably represented within the high-leverage subset, so that additional uniformly sampled data provide limited improvement for these observables.

A clearer distinction emerges in the dilute Cu-in-Al regime (Table~\ref{tab:alcu_dilute_aggregate}). At 50\% data, vacancy formation RMSEs range from $0.20$ to $0.24$\,eV for the heuristic selection strategies. In contrast, CUR at 25\% reduces the vacancy RMSE to $0.086$\,eV, substantially below the 50\% heuristic baselines, while block-CUR remains within the same general range as the baseline methods. In the dilute limit, where a single solute perturbs an otherwise host-dominated coordination environment, accurate vacancy energetics require resolving subtle chemical asymmetries in local bonding. The improved performance of CUR with half the labeled data suggests that leverage-guided selection can preferentially retain configurations that constrain these chemically sensitive directions of the potential-energy surface. Vacancy--solute binding energies are reproduced within $0.04$--$0.07$\,eV across all methods, indicating that short-range interaction trends are comparatively insensitive to the choice of selection strategy.

Taken together, these results show that alloy defect energetics provide a more discriminating validation metric than bulk RMSE alone. While ordered compounds exhibit similar performance across selection strategies once sufficient chemical diversity is present, the dilute regime reveals a clear advantage for leverage-CUR. In this limit, improved defect accuracy can be achieved using approximately half the number of labeled structures, highlighting the role of statistically informed sampling in improving data efficiency for complex alloy systems.

\begin{table*}[t]
\centering
\caption{Ordered Al--Cu compounds: aggregate defect errors vs DFT (eV). 
Antisite RMSE is computed across Al$_2$Cu, Al$_3$Cu and AlCu antisite pairs. 
Vacancy RMSE is computed across all ordered vacancy cases (V$_\mathrm{Al}$ and V$_\mathrm{Cu}$).
Values represent the mean over five independent runs, digits in the parentheses the standard deviation in the trailing digits.}
\label{tab:alcu_ordered_aggregate}
\begin{tabular}{lccc}
\toprule
Method & \% Data & Antisite (eV) & Vacancy (eV) \\

\midrule
Random        & 50\% & $0.267(49)$ & $0.160(25)$ \\
Energy-based  & 50\% & $0.275(62)$ & $0.152(18)$ \\
Force-based   & 50\% & $0.216(38)$ & $0.184(57)$ \\
\midrule
CUR           & 25\% & $0.29(15)$  & $0.175(48)$ \\
block-CUR     & 25\% & $0.278(79)$ & $0.174(48)$ \\

\bottomrule
\end{tabular}
\end{table*}

\begin{table*}[t]
\centering
\caption{Dilute Cu in Al: aggregate errors vs DFT (eV). 
Vacancy binding energy RMSE is computed across first-, second-, and third-nearest neighbour vacancies. 
Vacancy formation energy RMSE is computed across host and solute-vacancy configurations.
Values represent the mean over five independent runs, digits in the parentheses the standard deviation in the trailing digits.
}
\label{tab:alcu_dilute_aggregate}
\begin{tabular}{lccc}
\toprule
Method & \% Data & Binding (eV) & Formation (eV) \\

\midrule
Random        & 50\% & $0.056(11)$ & $0.204(41)$ \\
Energy-based  & 50\% & $0.071(18)$ & $0.243(47)$ \\
Force-based   & 50\% & $0.064(8)$  & $0.214(36)$ \\
\midrule
CUR           & 25\% & $0.043(12)$ & $0.086(64)$ \\
block-CUR     & 25\% & $0.065(17)$ & $0.164(56)$ \\

\bottomrule
\end{tabular}
\end{table*}

\subsection{Data efficiency and practical gains}
\label{subsec:data_efficiency}

The purpose of this section is to translate the learning-curve behavior into an explicit reduction of DFT workload. In many MLIP workflows, candidate configurations are first generated and then fully labeled with DFT; any subsequent pruning or reweighting therefore occurs only after the dominant computational cost has already been incurred. This ``label--then--filter'' ordering does not reduce the number of DFT evaluations required for the initial pool.

Leverage-guided selection changes this ordering because it is determined by the geometry of the regression problem rather than by the quantum-mechanical labels. The regularized leverage scores depend only on the weighted ACE design matrix $\widetilde A = W A$ and the chosen prior (Sec.~\ref{subsec:leverage_scores}), and can therefore be computed directly from atomic geometries and the ACE basis, without access to DFT energies or forces. This enables a ``filter--then--label'' workflow: a large structural pool may be generated inexpensively (e.g.\ via \textsc{Assyst}), leverage scores evaluated in descriptor space, and DFT performed only for the selected subset. In the present study, the full DFT-labeled pools are retained solely to enable controlled comparisons between selection strategies; the leverage criterion itself remains label-free.

The learning curves for Al, Cu, and Al--Cu support this interpretation. For elemental Al, block-CUR reaches the plateau regime within uncertainty at $f_{\mathrm{lab}}\approx 0.30$--$0.40$, whereas random sampling requires $f_{\mathrm{lab}}\approx 0.75$--$0.80$ to attain comparable test errors (Fig.~\ref{fig:al_learning_curves_Al}). A similar early-saturation behavior is observed for Cu. Operationally, this corresponds to an approximate two- to threefold reduction in the number of DFT-labeled configurations required to reach the plateau regime for these elemental systems within the present linear ACE model class.

For Al--Cu alloys, the situation is more structure-dependent. In ordered intermetallic compounds, leverage-based subsets trained on 25\% of the pool achieve defect errors statistically comparable to 50\% baselines, indicating diminishing returns once the dominant chemically distinct environments are represented. In the dilute regime, however, leverage-guided selection reaches improved vacancy accuracy using half the labeled data, demonstrating that selective sampling can more efficiently constrain chemically asymmetric local environments.

From a Bayesian linear-regression perspective, this behavior is consistent with leverage-based sampling preferentially selecting configurations that expand the span of $\widetilde A$, thereby reducing posterior parameter uncertainty and shrinking the dominant eigenmodes of the covariance. In this sense, the observed compression factors reflect a geometry-driven approximation to optimal experimental design, achieved without explicit determinant optimization.

By contrast, energy- and force-based heuristics require DFT labels to rank configurations and therefore do not modify the fundamental cost structure; they remain ``label--then--filter'' baselines. Because they rank by scalar residual magnitude rather than by coverage of descriptor space, they can converge more slowly or exhibit greater variability, particularly at small labeled fractions. 

Overall, these results demonstrate that geometry-aware, label-free subsampling provides a practical and statistically grounded route to reducing DFT labeling requirements for linear ACE models trained on systematically generated structure pools, while retaining comparable accuracy within statistical uncertainty for the systems examined here.

\section{Conclusions and Outlook}
\label{section:Conclusion}

This work examined how principled subset selection can reduce the DFT labeling burden in the development of linear ACE interatomic potentials. Rather than modifying model architecture or regression schemes, we focused on the statistical structure of the training data and assessed whether descriptor-space–guided selection can improve data efficiency without compromising transferability.
Because of our diverse training data and unrelated test cases that span multiple applications, point defects, planar defects, and multiple bulk phases, these results clearly indicate that data efficiency and transferability of the models can be maintained \emph{simultaneously} and  \emph{do not} inherently conflict with each other.

Using systematically generated \textsc{Assyst} structure pools, we compared random, energy-based, and force-based heuristics with leverage-guided CUR and block-CUR sampling on a fixed ACE architecture. For elemental Al and Cu, leverage-based subsets reached the test-error saturation regime at substantially smaller labeled fractions than uniform random sampling. In Al, comparable energy and force accuracy was obtained at $f_{\mathrm{lab}}\approx 0.30$--$0.40$, whereas random sampling required $f_{\mathrm{lab}}\approx 0.75$--$0.80$, corresponding to an effective two- to threefold reduction in DFT evaluations for this model class and dataset.

The impact of subset selection becomes clearer in defect-sensitive validation tests. In ordered Al--Cu intermetallic compounds, all strategies yielded similar antisite and vacancy errors once sufficient chemical diversity was present, and leverage-guided subsets preserved defect-level fidelity despite a twofold reduction in labeled data (25\% vs 50\%). In the dilute Cu-in-Al limit, however, leverage-based selection achieved markedly lower vacancy formation errors while using approximately half as many labeled configurations, indicating improved efficiency in constraining chemically asymmetric local environments. These findings highlight that aggregate RMSE metrics alone do not fully characterize transferability; the geometry of the selected training set plays a central role in capturing broken-symmetry and low-coordination configurations.

From a methodological standpoint, the results are consistent with a Bayesian linear-regression interpretation in which leverage sampling preferentially expands the span of the weighted design matrix and reduces posterior parameter uncertainty. CUR-type selection can therefore be viewed as a geometry-driven approximation to optimal experimental design, realized without explicit determinant optimization or reliance on DFT labels.

Several extensions follow naturally. Assessing scalability to more chemically complex multi-component alloys will clarify the limits of descriptor-based subsampling. Coupling leverage-based selection with dynamical or active-learning workflows may further improve efficiency by focusing DFT effort on both physically encountered and informationally informative configurations. Incorporating task-specific observables into the selection criterion may additionally enable training sets tailored to particular materials design objectives.

Overall, these results demonstrate that training-set construction is not merely a preprocessing step but a controllable lever for reducing quantum-mechanical workload in linear ACE potential development. Geometry-aware, label-free subset selection provides a practical and statistically grounded strategy for improving data efficiency while retaining physically meaningful transferability.

\section{Code availability}
Code for subset selection and model fitting will be released in a public repository; a link will be added upon publication.

\section*{Acknowledgements}

This work was supported by the NSERC CREATE program “Net Zero for Materials and Manufacturing” (Net0MM).

This research was enabled in part by support provided by the BC DRI Group and the Digital Research Alliance of Canada (alliancecan.ca).

MP and JN acknowledge funding from the Deutsche Forschungsgemeinschaft (DFG, German Research Foundation) through the Collaborative Research Center 1394 (SFB 1394, No. 409476157).

Computations were performed on the UBC ARC Sockeye high-performance computing platform and computational resources provided through the Stewart Blusson Quantum Matter Institute at the University of British Columbia.

Finally the authors would like to thank C. Ortner for this constructive comments on this manuscript.

\bibliographystyle{apsrev4-1}
\bibliography{apssamp}

\end{document}